\documentclass[journal,compsoc]{IEEEtran}

\usepackage{diagbox}

\usepackage{bm}
\usepackage{bbm}
\usepackage{graphicx}
\usepackage{enumerate}
\usepackage{caption}
\usepackage{algorithm}
\usepackage{algorithmic}
\usepackage{color}
\usepackage{subfigure}
\usepackage{graphicx}
\usepackage{epstopdf}
\usepackage{stfloats}
\usepackage{float}
\usepackage{amsthm}
\usepackage{amsmath}
\usepackage{booktabs}
\usepackage{multirow}
\usepackage{amssymb}

\usepackage[square, comma, sort&compress, numbers]{natbib}

\usepackage{color}
\usepackage{pifont}

\newtheorem{defi}{Definition}
\newtheorem{thm}{Theorem}

\begin{document}

\title{On the differential privacy of dynamic location obfuscation with personalized error bounds}

\author{Shun~Zhang,
	Benfei~Duan,
	Zhili~Chen$^\ast$,
and~Hong~Zhong
\IEEEcompsocitemizethanks{\IEEEcompsocthanksitem S. Zhang, B. Duan and H. Zhong are with School of Computer Science and Technology, Anhui University, Hefei 230601, China.\protect\\
E-mail: szhang@ahu.edu.cn (S. Zhang), dbf97@stu.ahu.edu.cn (B. Duan), zhongh@ahu.edu.cn (H. Zhong)
\IEEEcompsocthanksitem Z. Chen is with Software Engineering Institute, East China Normal University, Shanghai 200062, China.\protect \\
E-mail: zhlchen@sei.ecnu.edu.cn
\IEEEcompsocthanksitem $^\ast$ Corresponding author (Zhili Chen)}
\thanks{Manuscript received Month xx, 2020; revised Month xx, 2020.}}

\markboth{Journal of \LaTeX\ Class Files,~Vol.~xx, No.~x, Month~20xx}%
{Shell \MakeLowercase{\textit{et al.}}: Bare Demo of IEEEtran.cls for Computer Society Journals}
%



\IEEEtitleabstractindextext{
\begin{abstract}
Geo-indistinguishability and expected inference error are two complementary notions for location privacy. The joint guarantee of differential privacy (indistinguishability) and distortion privacy (inference error) limits the information leakage. In this paper, we analyze the differential privacy of PIVE, a dynamic location obfuscation mechanism proposed by Yu, Liu and Pu (NDSS 2017), and show that PIVE fails to offer either of the privacy guarantees on adaptive Protection Location Sets (PLSs) as claimed.
Specifically, we demonstrate that different PLSs could intersect with one another
due to the defined search algorithm, and then different apriori locations in the same PLS could have different protection diameters. As a result, we can show that the proof of
local differential privacy for PIVE is problematic.
Besides, 
the condition introduced in PIVE is confirmed to be not sufficient for bounding expected inference errors in general,
which makes the user-defined
inference error threshold invalid.
To address these issues, we propose a couple of correction approaches, analyze theoretically their satisfied privacy characteristics and detail their respective merits and demerits.

\end{abstract}

}

\maketitle

\IEEEdisplaynontitleabstractindextext

%
\IEEEpeerreviewmaketitle

\section{Introduction}
\IEEEPARstart{G}{EO-INDISTINGUISHABILITY} \cite{ABC13} and expected inference error
are two statistical quantification based privacy
notions for location obfuscation which attracts much
attention in recent years due to the rapid development of location-based services (LBSs) and mobile Internet.

Since Shokri \cite{Sho15} designs a
joint optimization approach for integrating the two privacy notions, geo-indistinguishability and expected
inference error, there have been some more works proposing to combine them, e.g., \cite{OTP17,YLP17}. In particular,
Yu et al. \cite{YLP17} study their relationship and summarize the problems as follows. First, the two notions are complementary for protecting
location privacy against inference attacks. Currently most existing location obfuscation mechanisms support only one of the two privacy notions. Second,
it is important to develop a location
obfuscation mechanism to effectively integrate both the
privacy notions. Third, incorporating user's desired privacy preference,
like minimum inference error threshold, improves the
usability, allows adaptive noise
adjustment for geo-indistinguishability and satisfies customizable
privacy/utility requirement of mobile users that permits
personalized error bounds at different situations related to locations, time
and LBSs. This motivates them to develop PIVE, a two-phase dynamic differential location
privacy scheme for preserving both notions of location privacy
with personalized error bounds.

Unfortunately, Yu et al.'s core approach for generating probability distribution is faced with privacy
problem. The problem results in that their mechanism can not theoretically preserve Differential Privacy (DP) \cite{DMN06}, on the
protection location set. Moreover, their proposed condition is not enough to ensure the lower bound of inference error.
For this, we investigate new approaches with personalized and joint privacy guarantees. Through theoretical analysis and simulation
experiments, we demonstrate the privacy problems and make some discussions on two ways of correction.

The rest of the paper is organized as follows. Section 2 mainly
reviews the core phases of Yu et al.'s PIVE.
Section 3 points out its privacy problems and provides theoretical
analysis and typical examples. Section 4 discusses two solution approaches to efficiently and correctly determine
the probability distribution with experiment results. Section 5 reviews related work,
and Section 6 concludes this paper.

\section{Review of PIVE Framework}

Yu et al. argue in \cite{YLP17} that geo-indistinguishability and expected
inference error are two complementary notions for location
privacy. They study the relationship and propose PIVE framework combining both notions.

\subsection{System Model}\label{subsec:model}

In PIVE, the authors consider a scenario where the user wants to protect the privacy of her/his actual location by reporting a pseudo-location
in a finite set $\mathcal{X}$ of nearby discrete locations.
It is desirable to develop a location obfuscation mechanism that combines the two privacy notions and generates perturbed locations with effective performance. The mechanism allows that the informed adversary has prior knowledge of distribution $\pi$ over $\mathcal{X}$ and knows obfuscation mechanism $f$, i.e., a probability distribution matrix.

The  objectives include: 1) preserve local Differential Privacy (DP) with respect to geo-indistinguishability, 2) protect location
against inference attacks of Bayesian adversary with
prior information, and 3) support customizable preferences for users on two privacy control knobs, DP parameter $\epsilon$ and minimum inference error bound $E_m$.


An obfuscation mechanism takes the actual location $x$ from $A$ as input and chooses a
pseudo-location $x^\prime$ from $O$ by sampling from the distribution $f(x^\prime|x)$:
\begin{equation}
f(x^\prime|x)=\text{Pr}(O=x'|A=x) \ \ \ \ \ x,\ x'\in \mathcal{X}.
\end{equation}
To quantify the adversary's absolute information gain, basic assumptions
on the prior knowledge are required \cite{Sho15,STT12}.
In the adversary model, the prior distribution $\pi$ over $\mathcal{X}$ is given in many ways such as the population density.
Based on the observed pseudo-location $x'$, the adversary computes the posterior probability distribution $\text{Pr}(x|x')$ that is the probability of $x$ to be the true location generating $x'$:
\begin{equation}\label{formu:post-dist}
\text{Pr}(x|x')=\frac{\text{Pr}(x,x')}{\text{Pr}(x')}=\frac{\pi(x)f(x'|x)}{\sum_{y\in \mathcal{X}}\pi(y)f(x'|y)}.
\end{equation}
Then a Bayesian adversary launches optimal inference attack usually by estimating the location $\hat x$ satisfying
\begin{equation}
\hat{x}=\mathop{\arg\min}\limits_{\hat{y}\in \mathcal{X}}\sum_{x\in \mathcal{X}}\text{Pr}(x|x')d_p(\hat{y},x),
\end{equation}
where $d_p$ denotes Euclidean distance $d$.

As a strong notion on location privacy, geo-indistinguishability is introduced in \cite{ABC13} due to an
extension of DP to arbitrary metrics \cite{CAB13}. It ensures that any two geographically close locations have similar probabilities to generate a pseudo-location. This allows that
 the actual location is protected by being hidden in a protection region, and the Protection
Location Set (PLS) consists of the locations within the protection region \cite{YLP17}. In order to achieve DP protection over PLS, we develop a loose definition as follows.

\begin{defi}[$(\epsilon_g,\theta)$-Geo-indistinguishability within PLS]\label{def:geoI}
Assume that a mechanism $\mathcal{A}$ satisfies, for any $x, y$ in PLS $\Phi\subset\mathcal{X}$,
\begin{equation}
\frac{f(x^\prime|x)}{f(x^\prime|y)}\leq  e^{\epsilon_g \left(d(x,y)+\theta\right)},\ \ \ \  x^\prime\in \mathcal{X},
\end{equation}
then $\mathcal{A}$ is $(\epsilon_g,\theta)$-geo-indistinguishable on $\Phi$. If $\theta=0$, we say that $\mathcal{A}$ gives $\epsilon_g$-geo-indistinguishability on $\Phi$ without deviation.
\end{defi}


\begin{defi}[Local DP on a Location Set \cite{YLP17}]\label{def:eps_DPPLS}
A randomized location obfuscation mechanism
$\mathcal{A}$ satisfies $\epsilon$-DP on the set $\Phi$ contained in $\mathcal{X}$, if for any $x , y \in\Phi$ and any output $x'\in\mathcal{X}$,
\begin{equation}\label{defi:eps-DP}
\frac{f(x^\prime|x)}{f(x^\prime|y)}
\leq  e^{\epsilon}.
\end{equation}
\end{defi}

Since the local DP is currently achieved via geo-indistinguishability (with operating on a single dataset $\mathcal{X}$),
we use the notion $\epsilon$-DP instead of $\epsilon$-LDP as in previous literatures \cite{Sho15,WYH19}. This paper investigates geo-indistinguishability and local DP on each PLS and for more general cases, even on the whole $\mathcal{X}$.
Accordingly for the neighborhood relationship,
two locations are regarded to be neighboring to each other if in the same PLS. Besides, as a quite
general mechanism preserving DP, the exponential mechanism requires a scoring function $q:\ \Phi\times\mathcal{X}\rightarrow \mathbb{R}$ which assigns a real-valued score to each point-point pair, ideally such that each $x'\in\mathcal{X}$ with better utility receives a higher score for a given point $x\in\Phi$.

\begin{defi}[Sensitivity on PLS \cite{DMN06}]\label{def:Sensitivity}
Let $x_1, x_2$ be any pair of neighboring locations (in PLS $\Phi$) and $x'\in\mathcal{X}$. The sensitivity of the scoring function $q$ on $\Phi$ is given by, its maximal change,
\begin{equation}
\Delta q = \sup_{x_1,\, x_2,\, x'} \left| {q(x_1,x') - q(x_2,x')} \right|.
\end{equation}
\end{defi}

\begin{defi}[Exponential Mechanism on PLS \cite{MT07,DR14}]\label{def:Exponential}
Given a scoring function $q$ on $\Phi\times\mathcal{X}$,
 the exponential mechanism
$\mathcal{M}(x,q)$ outputs $x' \in \mathcal{X}$  with probability proportional to $\exp \left(\frac{\epsilon q(x,x')}{2\Delta q}\right)$.
\end{defi}

\subsection{Solution Scheme PIVE}\label{subsec:scheme}
Yu et al. develop PIVE, a two-phase dynamic differential location privacy framework. In Phase I, they introduce user-defined inference error bounds $E_m$ to determine the PLS. In Phase II, they
produce pseudo-locations by deploying exponential mechanism using the Euclidean distance.


First, considering the optimal inference attack of Bayesian adversary,
 the conditional expected
inference error for any observed pseudo-location $x'$ is given by
\begin{equation}
ExpEr(x')=\mathop{\min}\limits_{\hat{x}\in \mathcal{X}}\sum_{x\in \mathcal{X}}\text{Pr}(x|x')d(\hat{x},x), \ \ \text{for}\ x'\in \mathcal{X},
\end{equation}
where the metric $d$ denotes still the Euclidean distance. Naturally, the global location privacy is usually measured by (unconditional) expected inference error,
\begin{equation}
\begin{split}
 ExpErr&=\sum_{x'\in \mathcal{X}} \text{Pr}(x') \cdot ExpEr(x')\\
 &=\sum_{x'\in \mathcal{X}}
    \mathop{\min}\limits_{\hat{x}\in \mathcal{X}}\sum_{x\in \mathcal{X}} \text{Pr}(x') \text{Pr}(x|x')d(\hat{x},x)
 \\
&=\sum_{x'\in \mathcal{X}}\mathop{\min}\limits_{\hat{x}\in \mathcal{X}}\sum_{x\in \mathcal{X}}\pi (x)f(x'|x)d(\hat{x},x).
\end{split}
\end{equation}

The service quality loss is given by the unconditional expected distance between true and perturbed locations,
\begin{equation}
QLoss=\sum_{x\in \mathcal{X}}\sum_{x'\in \mathcal{X}}\pi (x)f(x'|x)d(x',x).
\end{equation}

To guarantee $ExpEr(x')$ in terms of PLS,
the authors assume that the adversary narrows possible
guesses to the PLS that contains the user's true location. Afterwards, by normalization in PLS they
obtain the lower bound,
\begin{equation}\label{lower-bound}
\begin{split}
&DopEr(\Phi,x')=\mathop{\min}\limits_{\hat{x}\in \mathcal{X}}\sum_{x\in \Phi}  \frac{\text{Pr}(x|x')}{\sum_{y\in \Phi}\text{Pr}(y|x')}d(\hat{x},x)\\
=&\sum_{x\in \Phi}\frac{\text{Pr}(x|x')}{\sum_{y\in \Phi}\text{Pr}(y|x')}d(z',x) \quad\quad (\text{for some}\ \ z'\in \mathcal{X})\\
=& 
\sum_{x\in \Phi}\frac{\pi(x)f(x'|x)}{\sum_{y\in \Phi}\pi(y)f(x'|y)}d(z',x)\quad \quad(\text{due to}\  \eqref{formu:post-dist})\\
\ge & 
\sum_{x\in \Phi}\frac{\pi(x)f(x'|x)}{\sum_{y\in \Phi}\pi(y)e^{\epsilon}f(x'|x)}d(z',x)
\quad\quad (\text{due to}\  \eqref{defi:eps-DP})
\\
\ge & e^{-\epsilon}\mathop{\min}\limits_{\hat{x}\in \Phi}\sum_{x\in \Phi}\frac{\pi(x)}{\sum_{y\in \Phi}\pi(y)}d(\hat{x},x),
\end{split}
\end{equation}
\noindent where the last inequality holds true under the assumption that $\Phi$ is convex in $\mathcal{X}$.
Then they define
\begin{equation}
E(\Phi)=\mathop{\min}\limits_{\hat{x}\in \Phi}\sum_{x\in \Phi}\frac{\pi(x)}{\sum_{y\in \Phi}\pi(y)}d(\hat{x},x).
\end{equation}
This yields, together with \eqref{lower-bound},
\begin{equation}\label{ndss30}
ExpEr(x')\ge DopEr(\Phi,x')\ge e^{-\epsilon}E(\Phi).
\end{equation}
By \eqref{ndss30}, the authors in \cite{YLP17} obtain a sufficient condition,
\begin{equation}\label{ndss31}
E(\Phi)\ge e^{\epsilon} E_m,
\end{equation}
to satisfy the user-defined threshold, $ExpEr(x^\prime)\ge E_m$, for the inference attack using any observed pseudo-location $x'$.

\begin{figure}[tb]
\centering
\includegraphics[scale=0.35]{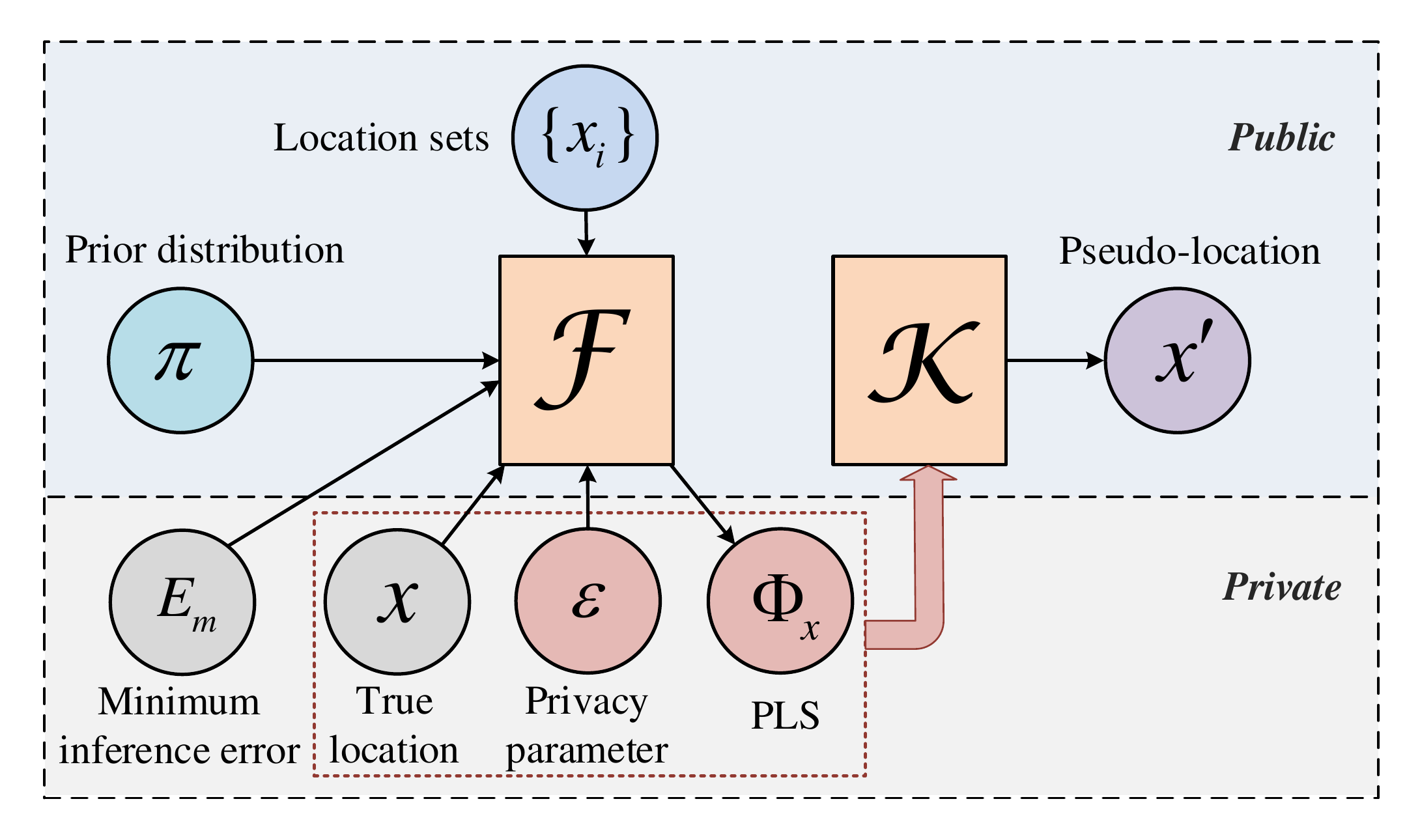}
\caption{The framework of PIVE.}
\label{fig:PIVE_model}
\end{figure}

We turn to the two phases of PIVE as shown in Fig. \ref{fig:PIVE_model}.

{\bf Phase I: Determining Protection Location Set.}
Based on \eqref{ndss31}, PIVE regards $\Phi$ as a variable and dynamically searches  region  $\Phi$ satisfying \eqref{ndss31} with diameter as small as possible.

\begin{figure}[tb]
\centering
	\subfigure[$4\times4$]{\label{fig:hilbert44}
		\includegraphics[scale=0.5]{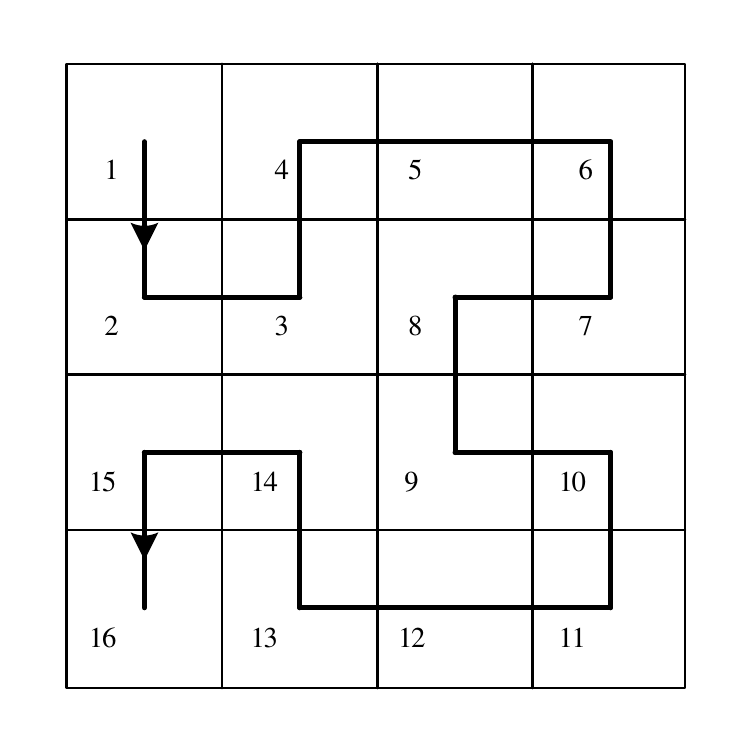}
	}
	\subfigure[$8\times8$]{\label{fig:hilbert88}
		\includegraphics[scale=0.5]{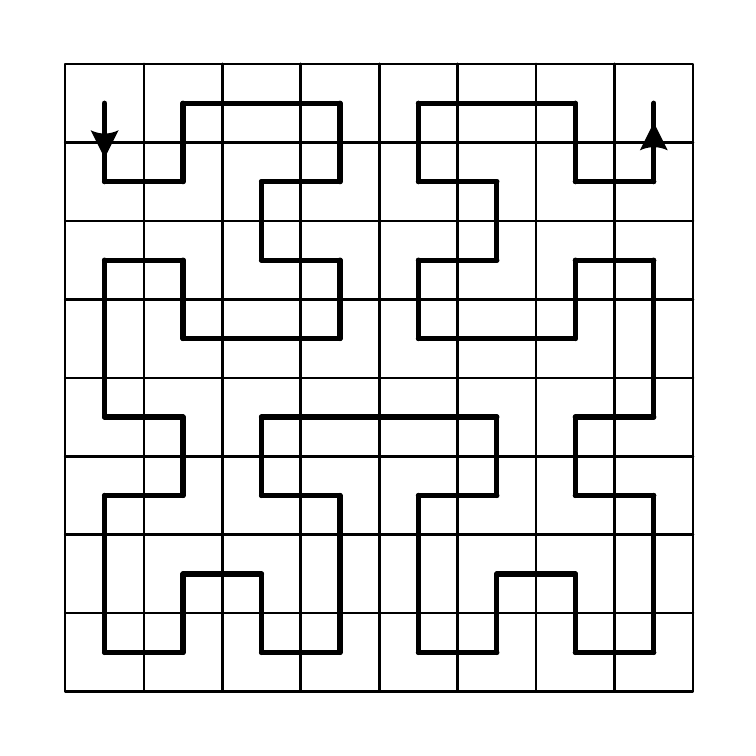}
	}
	\caption{Hilbert Curves for $4\times4$ and $8\times8$ grids.}
	\label{fig:hilbert}
\end{figure}

For conveniently determining PLS, the authors develop a Hilbert curve based search method. The Hilbert curve has the clustering properties with preserving geometrical adjacency \cite{MJF01}.
The domain can be covered by a $2^n \times 2^n$ grid that is naturally divided into $4$ rectangular
 parts each of which consists of a $2^{n-1} \times 2^{n-1}$ grid. The curve passes through $4$ parts clockwise or counterclockwise. Recursively, the curve passes within each part in the same manner.
It provides a
mapping from data point $x$ in a $2$-D space onto an $1$-D space with a so-called Hilbert value denoted by $H(x)$, see Fig. \ref{fig:hilbert} for two grids.
All locations in $\mathcal{X}$ are sorted according to their Hilbert values, and the rank of $x$ is denoted by $R(x)$ successively in $\mathcal{X}$, as shown in Fig. \ref{fig:Geolife_hilbert}.
The searching range only covers the
locations with ranks in $[R(x)- range, R(x) + range]$ over the sorted sequence of $\mathcal{X}$, where $range$ is a positive integer decided heuristically.

 {\bf Algorithm 1} is proposed in \cite{YLP17} to check
all possible sets $\{x_{-l},\ldots,x_0=x,x_1,\ldots,x_{r}\}$ (including actual location $x$) covered in searching range. It returns a set $\Phi_x$ having the smallest diameter
among those satisfying \eqref{ndss31}. Then
different locations may have different PLSs and diameters.

{\bf Phase II: Differentially Private Mechanism.}
The exponential mechanism
is devised to generate pseudo-locations, which
achieves local DP on the PLS and makes strong utility guarantee with respect to the service quality.

Since smaller distance produces higher utility then the utility (i.e., scoring) function is defined by $u_x(x')=-d(x,x')$, using the Euclidean distance between perturbed and true locations. The sensitivity on PLS is its diameter. The designed mechanism $\mathcal{K}$ outputs each $x^\prime$ from $\mathcal{X}$ with
the probability $w_x \exp(\frac{-\epsilon d(x,x^\prime)}{2D(\Phi)})$, where
\begin{equation}\label{wx}
w_x = \left( \sum_{x^\prime \in \mathcal{X} } \exp\left(\frac{-\epsilon d(x,x^\prime)}{2D(\Phi)}\right) \right)^{-1}.
\end{equation}

Following this, the user's true location $x$ determines diameter $D(\Phi)$ of PLS $\Phi$ and the sensitivity $\Delta u=D(\Phi)$. 
The authors argue theoretically, their designed exponential mechanism $\mathcal{K}$ preserves $\epsilon$-DP on the PLS $\Phi(x)$.

\section{Privacy Problems of PIVE Framework}\label{sec: PIVE-prob}

This section mainly shows that \emph{the PIVE does not guarantee Differential Privacy (DP)}, due to the conflict between the generation of Protection Location Sets (PLSs) and the application of the exponential mechanism, and we confirm \emph{the failure of control on expected inference errors with threshold $E_m$}.

First, we find that the generation of PLSs in PIVE leads to the property stated in Observation 1.

\emph{\textbf{Observation 1}. The PLSs generated by PIVE may have personalized diameters and intersect with each other.}


The PIVE generates PLSs with Algorithm 1 in \cite{YLP17} as follows.  For each input location $x$ denoted by $x_0$ (regarded as an actual location), the algorithm returns a set having the smallest diameter satisfying \eqref{ndss31}. The locations in the output set are required to have consecutive rankings in $\mathcal{X}$ according to their mappings on a Hilbert curve.
This means that each (true) location $x$ has its own PLS $\Phi_x$ and the corresponding diameter $D(\Phi_x)$, and different (even neighboring) locations have different PLSs with different diameters and even have intersections on PLSs. This can be demonstrated by a simple example as follows.


\begin{figure}[tb]
\centering
\includegraphics[scale=0.505]{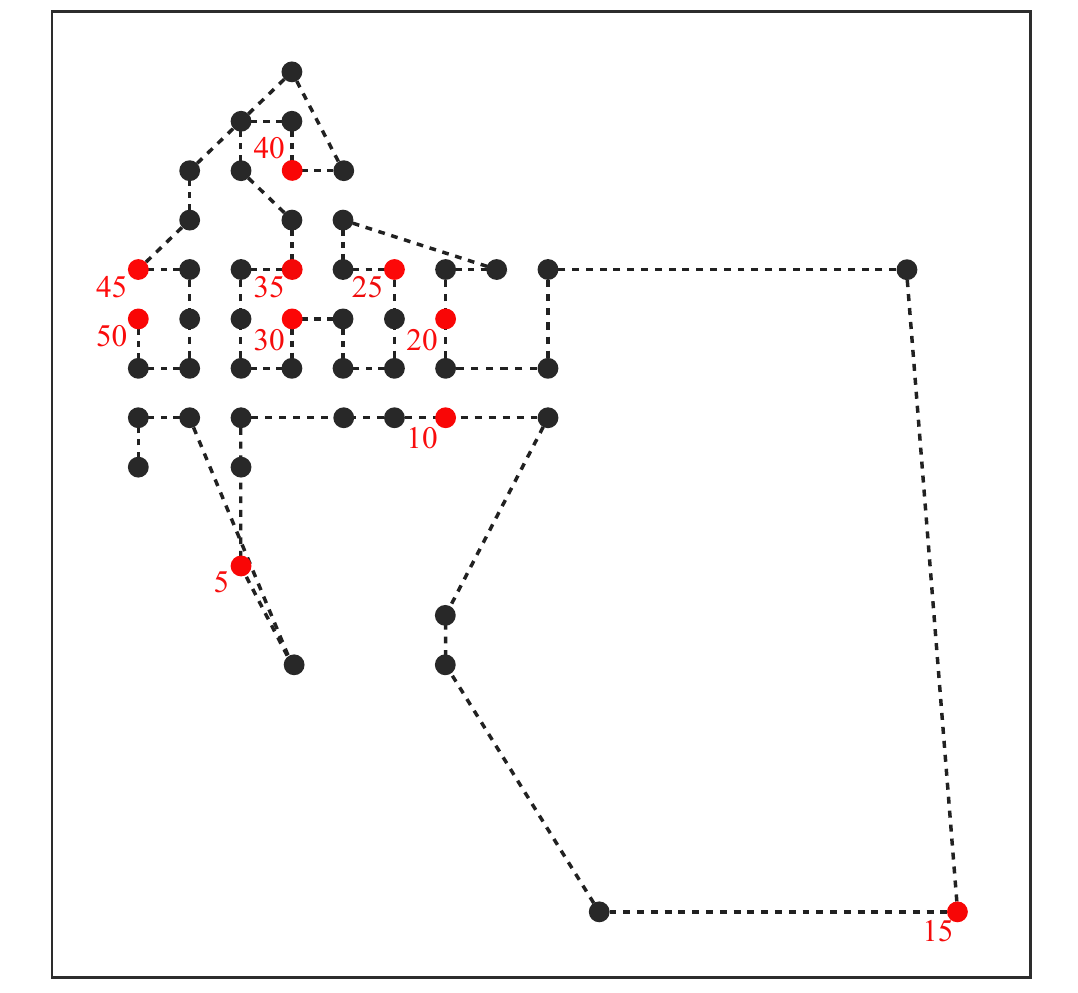}
\caption{50 regions in Geolife ranked on a Hilbert Curve.}
\label{fig:Geolife_hilbert}
\end{figure}

\begin{figure}[htbp]
\centering
\includegraphics[scale=0.37]{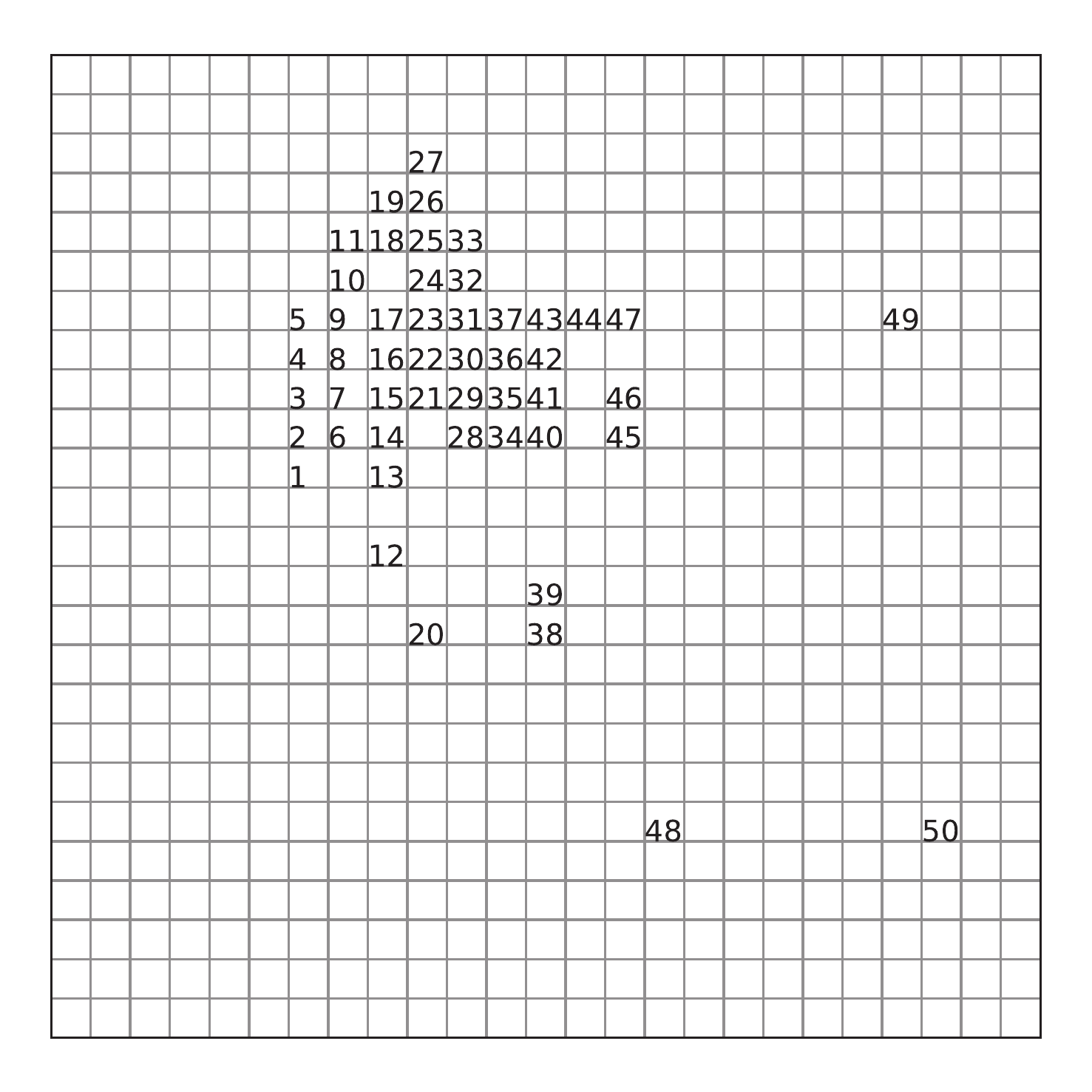}
\caption{50 regions in Geolife dataset.}
\label{fig:geolife}
\end{figure}

We implement PIVE with 50 regions in Geolife dataset distributed as in \cite{YLP17}, see Fig. \ref{fig:geolife}, For convenience, our grid is with the cell scale 1km$\times$1km and prior distribution $\pi$ values are uniformly and randomly generated in $[0.01,0.03]$,
 see \cite{ZDC21} for detailed experimental setup. In the default setting of $\epsilon=1.0$ and $E_m=0.15$, we execute Algorithm 1 in \cite{YLP17} to obtain PLSs with new rankings on a Hilbert curve as in Fig. \ref{fig:Geolife_hilbert}.  In Table \ref{tlb:pls},  with the rankings of locations improved by the above Hilbert curve, each number in bold means the initial point (user's actual location) of a PLS and the corresponding diameter for each PLS is given in the bracket behind.

\begin{table}[htbp]\renewcommand{\arraystretch}{1.5}
	\centering
    \scriptsize
	\caption{PLS for each location numbered on Hilbert Curve.}\label{tlb:pls}
	\begin{tabular}{|l|l|l|l|l|}
    \hline

    \textbf{1},2,3 (1.4)&\textbf{2},3 (1.0)&\textbf{3},2 (1.0)&\textbf{4},5 (2.2)&\textbf{5},6 (2.0)\\\hline
    \textbf{6},7 (1.0)&\textbf{7},6 (1.0)&\textbf{8},9 (1.0)&\textbf{9},10 (1.0)&\textbf{10},9 (1.0)\\\hline
    \textbf{11},10 (2.0)&\textbf{12},11 (4.5)&\textbf{13},11,12 (5.4)&\textbf{14},13 (5.8)&\textbf{15},14 (7.0)\\\hline
    \textbf{16},17 (7.0)&\textbf{17},18 (2.0)&\textbf{18},19 (2.0)&\textbf{19},20,21 (2.0)&\textbf{20},21,22 (1.4)\\\hline
    \textbf{21},20,22 (1.4)&\textbf{22},20,21 (1.4)&\textbf{23},24 (1.0)&\textbf{24},23 (1.0)&\textbf{25},24,26 (1.4)\\\hline
    \textbf{26},27 (1.0)&\textbf{27},28 (1.0)&\textbf{28},29 (1.0)&\textbf{29},30 (1.0)&\textbf{30},31 (1.0)\\\hline
    \textbf{31},30 (1.0)&\textbf{32},31,33 (1.4)&\textbf{33},34 (1.0)&\textbf{34},35 (1.0)&\textbf{35},34 (1.0)\\\hline
    \textbf{36},37 (1.4)&\textbf{37},38 (1.0)&\textbf{38},37 (1.0)&\textbf{39},40 (1.0)&\textbf{40},41 (1.0)\\\hline
    \textbf{41},40 (1.0)&\textbf{42},41 (2.2)&\textbf{43},44 (1.0)&\textbf{44},43 (1.0)&\textbf{45},46 (1.0)\\\hline
    \textbf{46},47 (1.0)&\textbf{47},48 (1.0)&\textbf{48},47 (1.0)&\textbf{49},50 (1.0)&\textbf{50},49 (1.0)\\\hline
	\end{tabular}
\end{table}

\begin{table}[htbp]\renewcommand{\arraystretch}{1.5}
	\centering
\scriptsize
	\caption{Prior probability distribution on related locations.}\label{tlb:pi}
	\begin{tabular}{|c|c|c|c|}
    \hline
    \textbf{region}&5&6&7\\\hline
    \bm{$\pi$}&0.0224&0.0153&0.0150\\\hline
	\end{tabular}
\end{table}

\begin{table}[htbp]\renewcommand{\arraystretch}{1.5}
	\centering
\scriptsize
	\caption{Computing $E(\Phi)$ for related sets.}\label{tlb:E_phi}
	\begin{tabular}{|c|c|c|c|}
    \hline
    \bm{$\Phi$}&\{\textbf{5},6\}&\{\textbf{6},7\}\\\hline
    \bm{$E(\Phi)$}&0.81&0.50\\\hline
	\end{tabular}
\end{table}

\begin{figure}[htbp]
\centering
\includegraphics[scale=0.35]{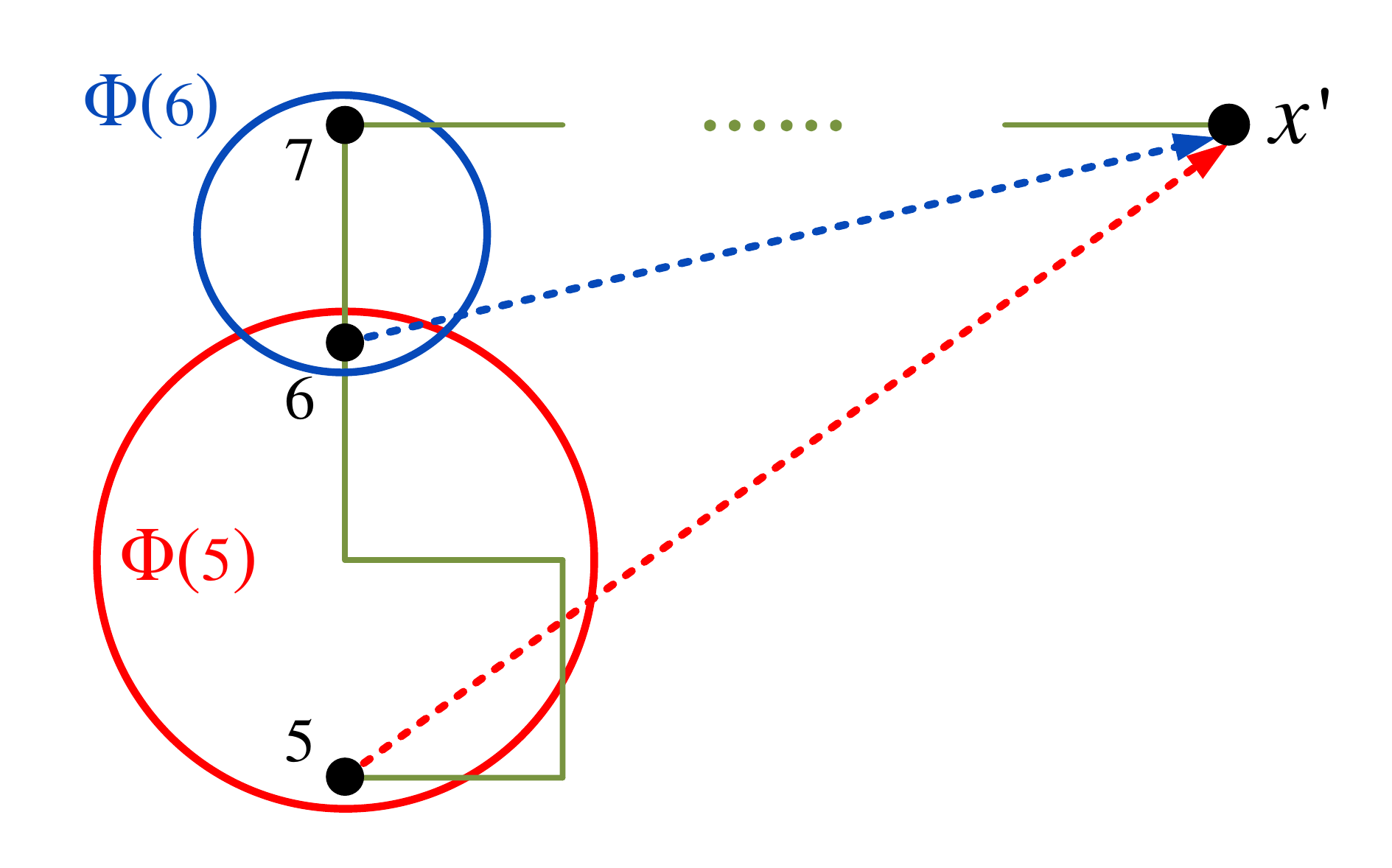}
\caption{Adaptive PLSs with intersections.}
\label{fig:PIVE_fig1}
\end{figure}

Specifically, the locations ${5}$ and ${6}$ have PLSs $\Phi(5)$ and $\Phi(6)$, respectively, see Fig. \ref{fig:PIVE_fig1}.
Related prior distribution $\pi$ are shown in Table \ref{tlb:pi} and the computations of some necessary $E(\Phi)$ are given in Table \ref{tlb:E_phi} with $e^{\epsilon}E_m=0.408$.
  We mention that the set $\{{\bf 5},6\}$ is preferred to $\{{\bf 5},4\}$ due to smaller diameter.
The location ${6}$ is included in PLS $\Phi(5)$ (with diameter $2.0$) and its corresponding PLS $\Phi(6)$ has smaller diameter $1.0$.
This demonstrates our Observation 1 and more intersections are shown in Table \ref{tlb:pls}.

Second, we have Observation 2 due to Observation 1 and the perturbation solution through exponential mechanism.

\emph{\textbf{Observation 2}. Different locations in the same PLS may have different diameters for applying the exponential mechanism.}

Following the example given by Fig. \ref{fig:PIVE_fig1}, we obtain that the PLSs $\Phi(5)$ and $\Phi(6)$ intersects with each other and have different diameters although both locations 5 and 6 are included in $\Phi(5)$.
Further, in Phase II, different rows represent different input locations and often have different diameters of PLSs, i.e., different sensitivities for achieving personalized preferences.




Following this, we show that the proof of DP guarantee in \cite{YLP17} is problematic. In particular, the diameters of locations $x$ and $y$ in a given PLS can not be regarded as the same due to the Observation 2.
That is, in the proof of differential privacy (Theorem 3 in \cite{YLP17}), i.e.,
\begin{equation}\label{f-over-f}
\frac{f(x^\prime|x)}{f(x^\prime|y)}=
\frac{w_x \exp\big( -\epsilon d(x,x^\prime)/\left( 2D(\Phi_x) \right) \big)}{w_y \exp\big( -\epsilon d(y,x^\prime)/\left( 2D(\Phi_y) \right) \big)},
\end{equation}
We have to mention that $D(\Phi_x) \neq D(\Phi_y)$ holds true in general. Then we can not simply use the
 triangular inequality, $|d(x,x^\prime)-d(y,x^\prime)|\le d(x,y)$, in \eqref{f-over-f} for proving the preservation of differential privacy as desired in \cite{YLP17}.

Indeed, in the PLS, $\Phi_t$ determined by a true location $t$, any two points $x,y \in \Phi_t$ have their own PLSs, $\Phi_x$ and $\Phi_y$, respectively, due to Algorithm 1 in \cite{YLP17} for optimal searching. We should regard $x$ and $y$ as true locations in the computation of $f(x^\prime|x)$ and $f(x^\prime|y)$, respectively. Then we have to use $D(\Phi_x)$ and $D(\Phi_y)$ as sensitivity for their PLSs, respectively. $\Phi_t$ is certainly a candidate PLS for the location $x$ but not the optimal one in general since $D(\Phi_x)\le D(\Phi_t)$. The same statements hold also for the location $y$.


Finally, we mention Observation 3 about lower bound of inference errors for achieving a personalized threshold.

\emph{\textbf{Observation 3}. The condition \eqref{ndss31} is not sufficient for guaranteeing the minimum inference error $E_m$.}

As for PLSs, the PIVE includes two doubtful claims: 1) $ExpEr(x')$ has the lower bound $DopEr(\Phi,x')$ by narrowing guesses within the actual PLS while the ``narrow'' assumption is in contradiction with the \emph{private} PLS, and 2) $DopEr(\Phi,x')\ge e^{-\epsilon}E(\Phi)$ for convex $\Phi$. Both are false in general, which rejects the sufficient condition \eqref{ndss31}.

Indeed, the first claim holds only in special cases, 
for instance if the outside locations are far from the PLS.
In the default setting, $DopEr(\Phi,x^\prime)>ExpEr(x^\prime)$ holds with probability $12.43\%$ in  global sense. 
In essence, the second claim deploys that $E'(\Phi) = E(\Phi)$ for convex $\Phi$, where

\vspace{-2mm}
\begin{equation}\label{def:E-prime}
E'(\Phi)=\mathop{\min}\limits_{\hat{x}\in \mathcal{X}}\sum_{x\in \Phi}\frac{\pi(x)}{\sum_{y\in \Phi}\pi(y)}d(\hat{x},x).
\end{equation}
An example in \cite{ZDC21} shows that $E'(\Phi) < E(\Phi)$ for some convex $\Phi$. The PLSs from PIVE are not convex in general.

Following the line of \eqref{lower-bound}, we have for any $\Phi$ with $\epsilon$-DP,
\begin{equation}\label{ndss30-pro}
PivEr(\Phi,x')\geq e^{-\epsilon}E(\Phi),
\end{equation}
 where the local inference error, no less than $DopEr(\Phi,x')$,
 \begin{equation}\label{local-Em}
PivEr(\Phi,x')=\mathop{\min}\limits_{\hat{x}\in \Phi}\sum_{x\in \Phi}  \frac{\text{Pr}(x|x')}{\sum_{y\in \Phi}\text{Pr}(y|x')}d(\hat{x},x),
\end{equation}
is achieved when an adversary narrows guesses \emph{unfairly} to the private $\Phi$'s including the true location $x$.
Hence,
condition \eqref{ndss31} for each PLS in PIVE
ensures not $ExpEr(x^\prime)\ge E_m$ but $PivEr(x^\prime)= \min_{y\in\mathcal{X}} PivEr(\Phi_y,x^\prime)\ge E_m$.  

\section{Two Possible Corrections}

To solve the privacy problems pointed out above,
the approach should ensure:
1) the perturbation matrix $\{f(x_j|x_i)\}$ (probability distribution) is generated before the input of user's true location, and
2) each location in the same Protection Location Set (PLS) shares an identical sensitivity $D$.

Following the above requirements, all locations in each PLS are regarded symmetrically as the possible true location.
 The change of true location should not result in any variation of the public probability distribution matrix $f$. Otherwise, it possibly causes some attacks by analyzing the change of the public probability distribution.

We investigate possibly feasible approaches in two directions, on the aspect whether PLSs are allowed to intersect with each other or not. Particularly, we address the issues on geo-indistinguishability and local Differential Privacy (DP) within each PLS and for more general cases.

\subsection{Uniform Sensitivity Approach}\label{subsec:uniform}

The uniform sensitivity approach allows that PLSs intersect with each other while replacing the bound $ExpEr(x^\prime)\ge E_m$ on $\mathcal{X}$ by $PivEr(x^\prime)\ge \widetilde E_m$ over PLSs. For this local bound, it also assumes the bad case that the adversary narrows
guesses to some private PLS covering  actual location.

Such a local inference error threshold is uniform for all locations and 
can limit an adversary's capability of distinguishing locations (the expected privacy level) within each PLS $\Phi_y$ that covers the actual $x$ 
while DP and local DP measure the privacy level in the worst case.
Due to \eqref{ndss30-pro},
\begin{equation}\label{ndss31-pro}
E(\Phi)\ge e^{\epsilon} \widetilde E_m,\quad \text{for each PLS}\ \Phi,
\end{equation}
implies $PivEr(x^\prime)\ge \widetilde E_m$ for the optimal inference attack.

 Given the privacy control knobs, $\widetilde E_m$ and $\epsilon$, this approach finds a dynamic PLS $\Phi(x_i)$ satisfying \eqref{ndss31-pro} for each $x_i$ as PIVE proceeds actually.
Then it defines the sensitivity uniformly as the maximum diameter $D_{\rm max}=\max_{i}\{D(\Phi(x_i))\}$.
Provided a true location $x$, it publishes a pseudo-location $x'\in \mathcal{X}$, by the exponential mechanism with the probability,
\begin{equation}\label{equ:uniform}
f(x'|x) = w_x \exp\left(\frac{-\epsilon d(x,x^\prime)}{2D_{\rm max}}\right),
\end{equation}
where
\begin{equation}\label{equ:uniformwx}
w_x = \left( \sum_{x^\prime \in \mathcal{X} } \exp\left(\frac{-\epsilon d(x,x^\prime)}{2D_{\rm max}}\right) \right)^{-1}.
\end{equation}

\begin{figure}[tb]
\centering
\includegraphics[scale=0.37]{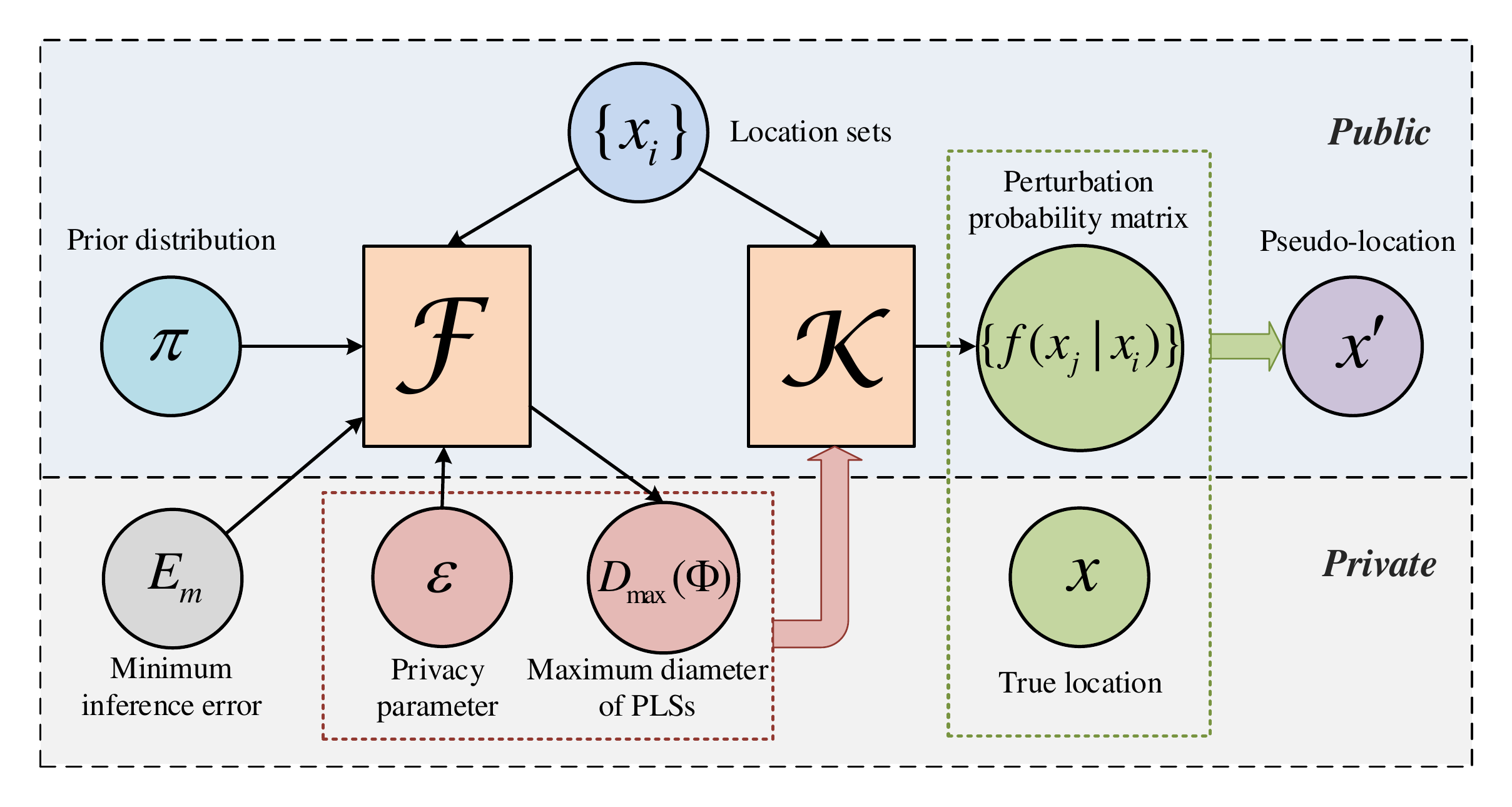}
\caption{Framework of uniform sensitivity approach.}
\label{fig:Dmax_PIVE}
\end{figure}
In this approach, although the publication diameters of all PLSs are extended to $D_{\rm max}$, their ranges remain unchanged as PIVE, since an extended PLS may not satisfy the condition \eqref{ndss31-pro}. The procedure is shown in Fig. \ref{fig:Dmax_PIVE}.

\begin{figure}[htbp]
\centering
	\subfigure[Uniform Sensitivity]{
		\includegraphics[scale=0.34]{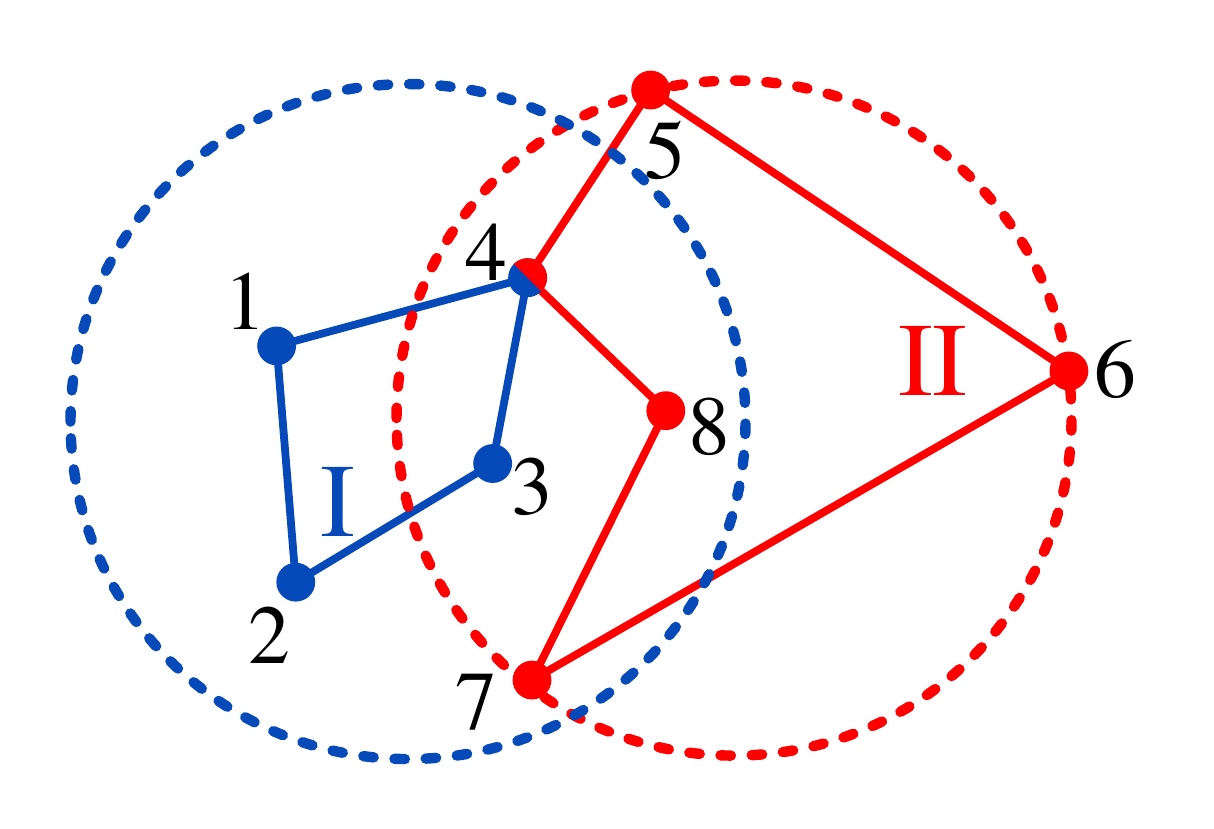}\label{fig:Dmax_simu}
	}
	\subfigure[Personalized Sensitivity]{
		\includegraphics[scale=0.34]{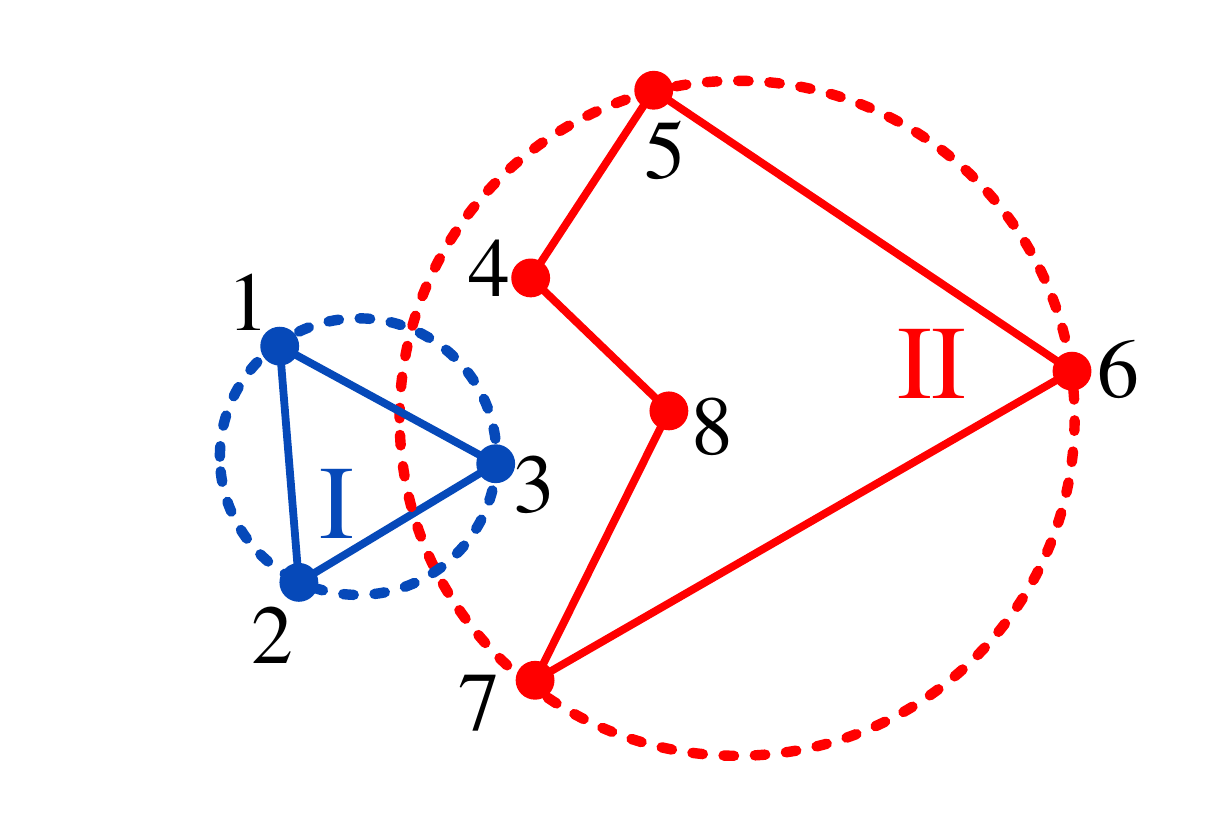}\label{fig:DPIVE_simu}
	}
	\caption{Examples of both correction approachs.}
\end{figure}

A simple example including $8$ locations, as shown in Fig. \ref{fig:Dmax_simu}, demonstrates the uniform sensitivity approach. There are PLS I (locations $1$-$4$) and PLS II (locations $4$-$8$), which intersect on location $4$. Obviously, the PLS II has the larger diameter denoted by $D_{\rm max}$, then the approach uses $D_{\rm max}$ as the sensitivity for both PLSs. The blue dotted circle is drawn with the same center of the circumcircle of PLS I. Both dotted circles describe only the mechanism sensitivity but not the ranges of PLSs. So, location $3$ is out of PLS II while locations $7$ and $8$ are out of PLS I. That is, the ranges of both PLSs do not change although their sensitivities may become larger.


\vspace{-3mm}

Following the above line, this uniform approach achieves $(\epsilon_g,D_{\rm max})$-geo-indistinguishability on the more general PLS that is the set of locations within any circular region with diameter of $D_{\rm max}$ in $\mathcal{X}$. For instance, PLS II in Fig. \ref{fig:Dmax_simu} can be extended by adding location $3$ with respect to indistinguishability. Theoretically,
\begin{equation}
\begin{split}\label{f-over-f-geo}
&\frac{f(x^\prime|x)}{f(x^\prime|y)}
=\frac{\exp \left(\frac{-\epsilon d(x,x^\prime)}{2D_{\rm max}}\right)}{\exp \left(\frac{-\epsilon d(y,x^\prime)}{2D_{\rm max}}\right)}
 \frac{\sum_{x^\prime \in \mathcal{X}}\exp \left(\frac{-\epsilon d(y,x^\prime)}{2D_{\rm max}}\right)}{\sum_{x^\prime \in \mathcal{X}}\exp \left(\frac{-\epsilon d(x,x^\prime)}{2D_{\rm max}}\right)}
\\
\leq & \exp\left(\frac{\epsilon |d(x,x^\prime)-d(y,x^\prime)|}{2D_{\rm max}}\right) \frac{\sum_{x^\prime \in \mathcal{X}}\exp \left(\frac{-\epsilon (d(x,x^\prime)-D_{\rm max})}{2D_{\rm max}}\right)}{\sum_{x^\prime \in \mathcal{X}}\exp \left(\frac{-\epsilon d(x,x^\prime)}{2D_{\rm max}}\right)}\\
\leq & e^{ d(x,y) \epsilon/(2D_{\rm max})} e^{{\epsilon}/{2}} \\
=&e^{\epsilon_g (d(x,y)+D_{\rm max})} \ \ \ ((\epsilon_g, D_{\rm max})\text{-geo-indistinguishability})\\
\leq& e^{2 \epsilon_g D_{\rm max}}=e^\epsilon \quad\ \ \, (\epsilon\text{-DP}),
\end{split}
\end{equation}
with $\epsilon_g=\epsilon/(2D_{\rm max})$.
Moreover, this approach using $D_{\max}$ preserves $\epsilon$-DP on each PLS $\Phi_x$, and on each more general PLS above if ignoring the local inference error threshold $\widetilde E_m$.

\begin{figure}[tb]
\centering
	\subfigure[Privacy V.S. $\epsilon$]{
		\includegraphics[height=1.25 in]{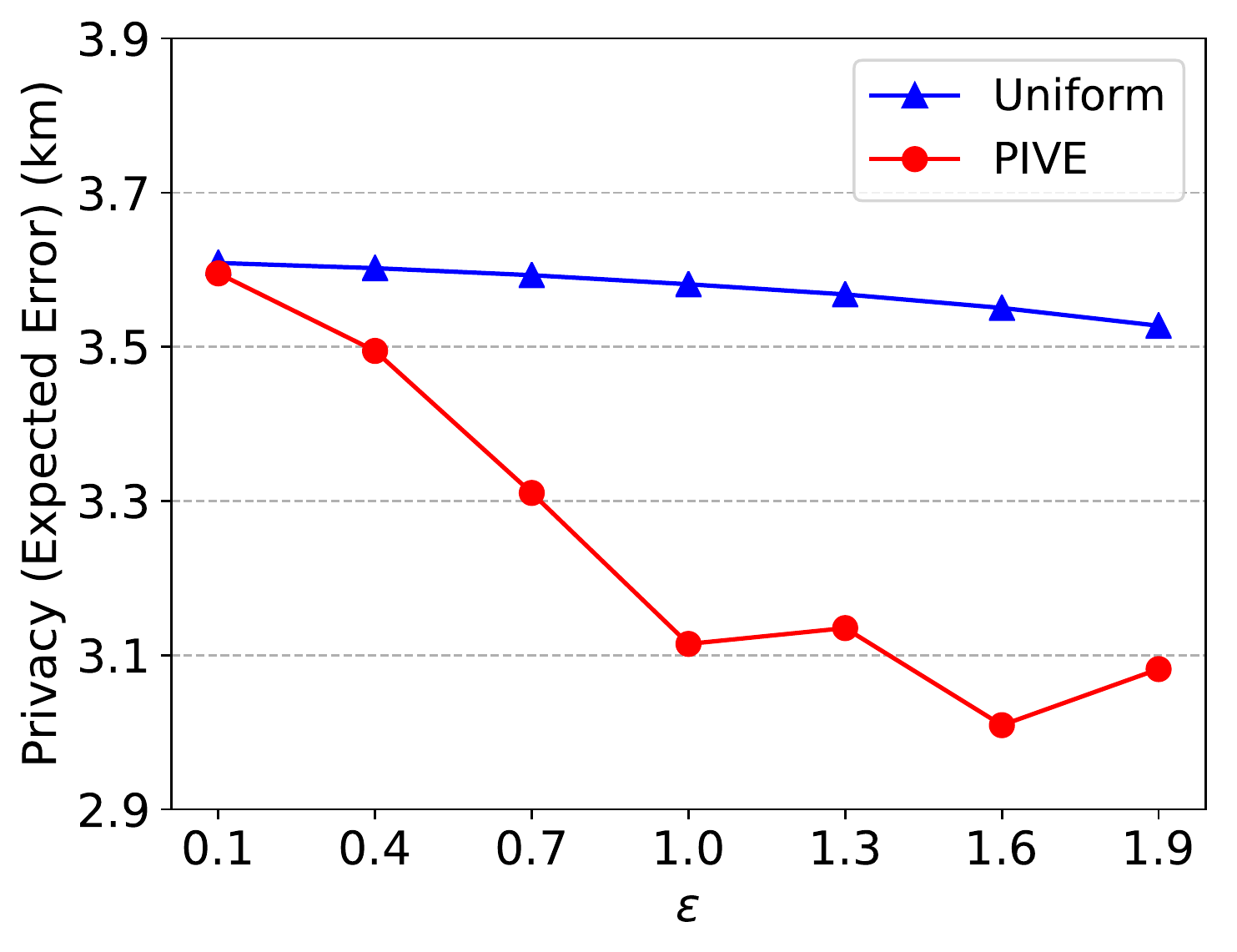}
	}
	\subfigure[Quality Loss V.S. $\epsilon$]{
		\includegraphics[height=1.25 in]{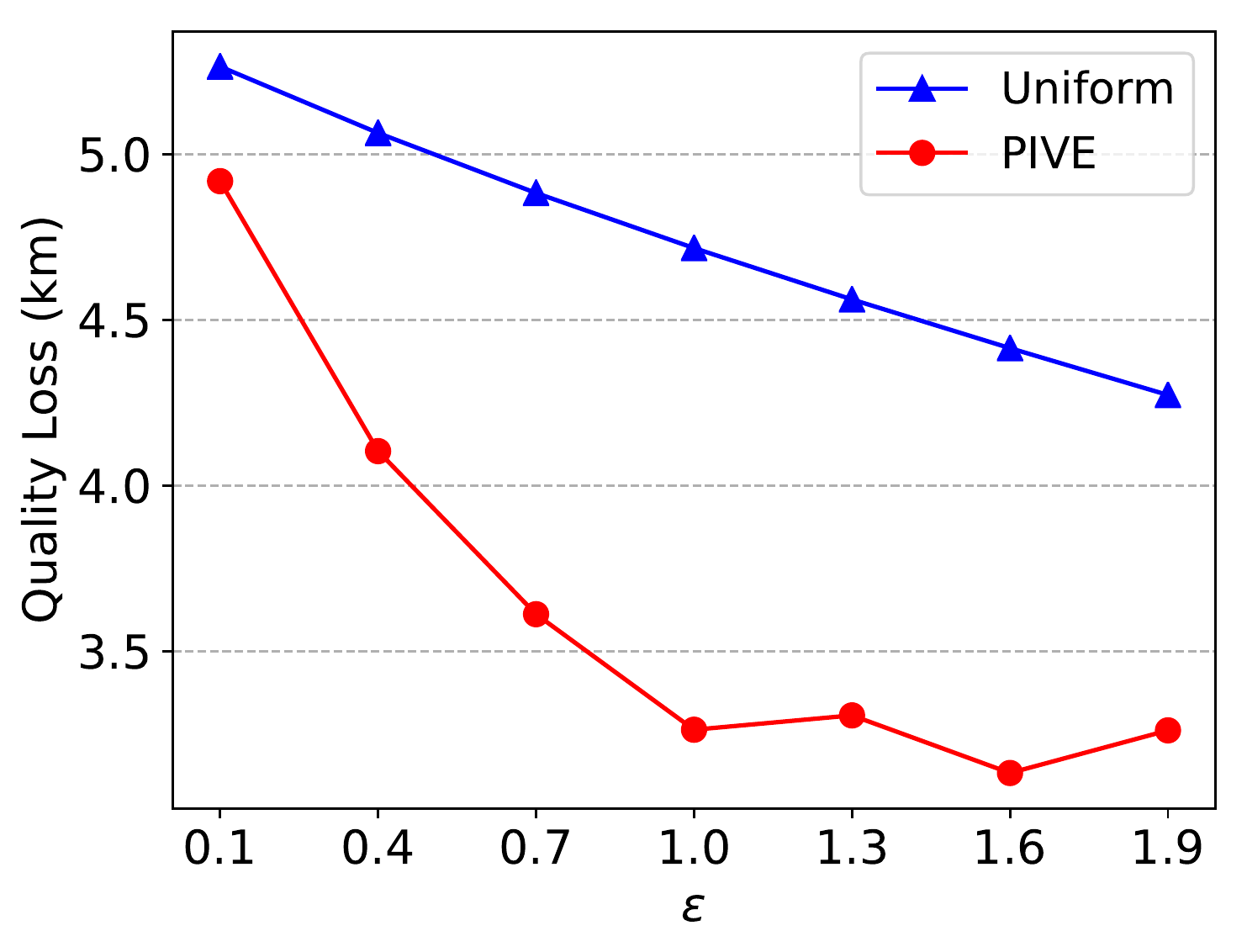}
	}
	\caption{Impact of $\epsilon$ under $D_{\rm max}$ strategy.}
	\label{fig:Dmax}
\end{figure}

However, the uniform sensitivity approch directly destroys the personalization of mechanism sensitivity for each apriori location.
In particular, when there exist obvious isolated locations, this approach could damage the utility of mechanism greatly. Following the experimental setting in Section \ref{sec: PIVE-prob} with $\widetilde E_m = E_m$, Fig. \ref{fig:Dmax} shows that the uniform sensitivity approach brings $29.5\%$ growth on average service quality loss for varied $\epsilon$, compared with the PIVE.

Finally, we conclude the features of the uniform sensitivity approach: 1) unable to achieve the personalization of sensitivity, then affecting the data utility largely, 2) unable to personalize error bound $\widetilde E_m$ for each location, 3)  $(\epsilon_g,D_{\rm max})$-geo-indistinguishable and $\epsilon$-DP uniformly on each PLS, 4) suitable for the scene of relatively uniformly distributed location domains without isolations, and 5) unavoidable to assume narrowing guesses to PLSs. Due to private PLSs, a personalized sensitivity approach is introduced without ``narrowing guesses'' as follows.

\subsection{Personalized Sensitivity Approach}\label{subsec:Personalized}

The personalized sensitivity approach requires that different PLSs do not intersect with one another, otherwise the exact privacy problem of PIVE arises. Then searching for PLSs becomes partitioning locations into groups $\{\Phi_k\}$. As a result, all locations in the same PLS share an identical sensitivity.



Currently we consider to transfer the global lower bound of inference error $ExpEr(x')$ to requirements on each PLS,
without the assumption that the adversary narrows 
guesses 
to the actual PLS. Let  $z=\mathop{\rm argmin}\limits_{\hat{x}\in \mathcal{X}}\sum_{x\in \mathcal{X}}\text{Pr}(x|x')d(\hat{x},x)$
 and denote $\text{Pr}(\Phi_k|x') = \sum_{y\in \Phi_k}\text{Pr}(y|x')$.
By normalization in each PLS $\Phi_k$ (with $\epsilon$-DP) from a partition $\{\Phi_k\}$,  we have
\begin{equation}
\begin{split}
&ExpEr(x')=\sum_{x\in \mathcal{X}}\text{Pr}(x|x')d(z,x)\\
=&\sum_{k}\sum_{x\in \Phi_k}\text{Pr}(x|x')d(z,x)\\
\ge &\sum_{k}\mathop{\min}\limits_{\widehat{x}_k\in \mathcal{X}}\sum_{x\in \Phi_k}\text{Pr}(x|x')d(\widehat{x}_k,x)\\
=& \sum_{k} \text{Pr}(\Phi_k|x') \mathop{\min}\limits_{\widehat{x}_k\in \mathcal{X}}\sum_{x\in \Phi_k}\frac{\text{Pr}(x|x')d(\widehat{x}_k,x)}{\sum_{y\in \Phi_k}\text{Pr}(y|x')}
 \\
=& 
\sum_{k} \text{Pr}(\Phi_k|x') \mathop{\min}\limits_{\widehat{x}_k\in \mathcal{X}}\sum_{x\in \Phi_k}\frac{\pi(x)f(x'|x)d(\widehat{x}_k,x)}{\sum_{y\in \Phi_k}\pi(y)f(x'|y)}
\\
\ge & 
\sum_{k} \text{Pr}(\Phi_k|x') \mathop{\min}\limits_{\widehat{x}_k\in \mathcal{X}}\sum_{x\in \Phi_k}\frac{\pi(x)f(x'|x)d(\widehat{x}_k,x)}{\sum_{y\in \Phi_k}\pi(y)e^{\epsilon}f(x'|x)}
\\
= & \sum_{k} \text{Pr}(\Phi_k|x') e^{-\epsilon} E'(\Phi_{k}), 
\end{split}
\end{equation}
\noindent
where for each $\Phi$ from a partition $\{\Phi_k\}$ of $\mathcal{X}$,

\begin{equation}
E'(\Phi)=\mathop{\min}\limits_{\hat{x}\in \mathcal{X}}\sum_{x\in \Phi}\frac{\pi(x)}{\sum_{y\in \Phi}\pi(y)}d(\hat{x},x).
\end{equation}
Since $\sum_{k} \text{Pr}(\Phi_k|x')=1$, 
the condition
that for all $\Phi_{k}$,
\begin{equation}\label{ndss31-pri}
E'(\Phi_k)\ge e^{\epsilon} E_m,
\end{equation}
implies the given threshold, $ExpEr(x^\prime)\ge E_m$. Formally,

\begin{thm}[]\label{thm:DPIVE}
Given a domain partition $\{\Phi_k\}$ and an observed pseudo-location $x^\prime$, suppose that an obfuscation mechanism satisfies $\epsilon$-DP on each PLS $\Phi_k$. If $E'(\Phi_k)\geq e^{\epsilon}E_m$ for each $\Phi_k$, then we have
 $ExpEr(x^\prime)\ge E_m$ for the optimal inference attack.
\end{thm}

For each PLS $\Phi$, this approach defines its sensitivity as its diameter $D(\Phi)$ for the exponential mechanism. Then, given a true location $x \in \Phi$, the approach publishes a pseudo-location $x'\in \mathcal{X}$ with the following probability,

\begin{equation}\label{equ:uniform}
f_{\Phi}(x'|x) = w^{\Phi}_x \exp\left(\frac{-\epsilon d(x,x^\prime)}{2D(\Phi)}\right),
\end{equation}
where

\begin{equation}\label{equ:uniformwx}
w^{\Phi}_x = \left( \sum_{x^\prime \in \mathcal{X} } \exp\left(\frac{-\epsilon d(x,x^\prime)}{2D(\Phi)}\right) \right)^{-1}.
\end{equation}

\begin{figure}[tb]
\centering
\includegraphics[scale=0.37]{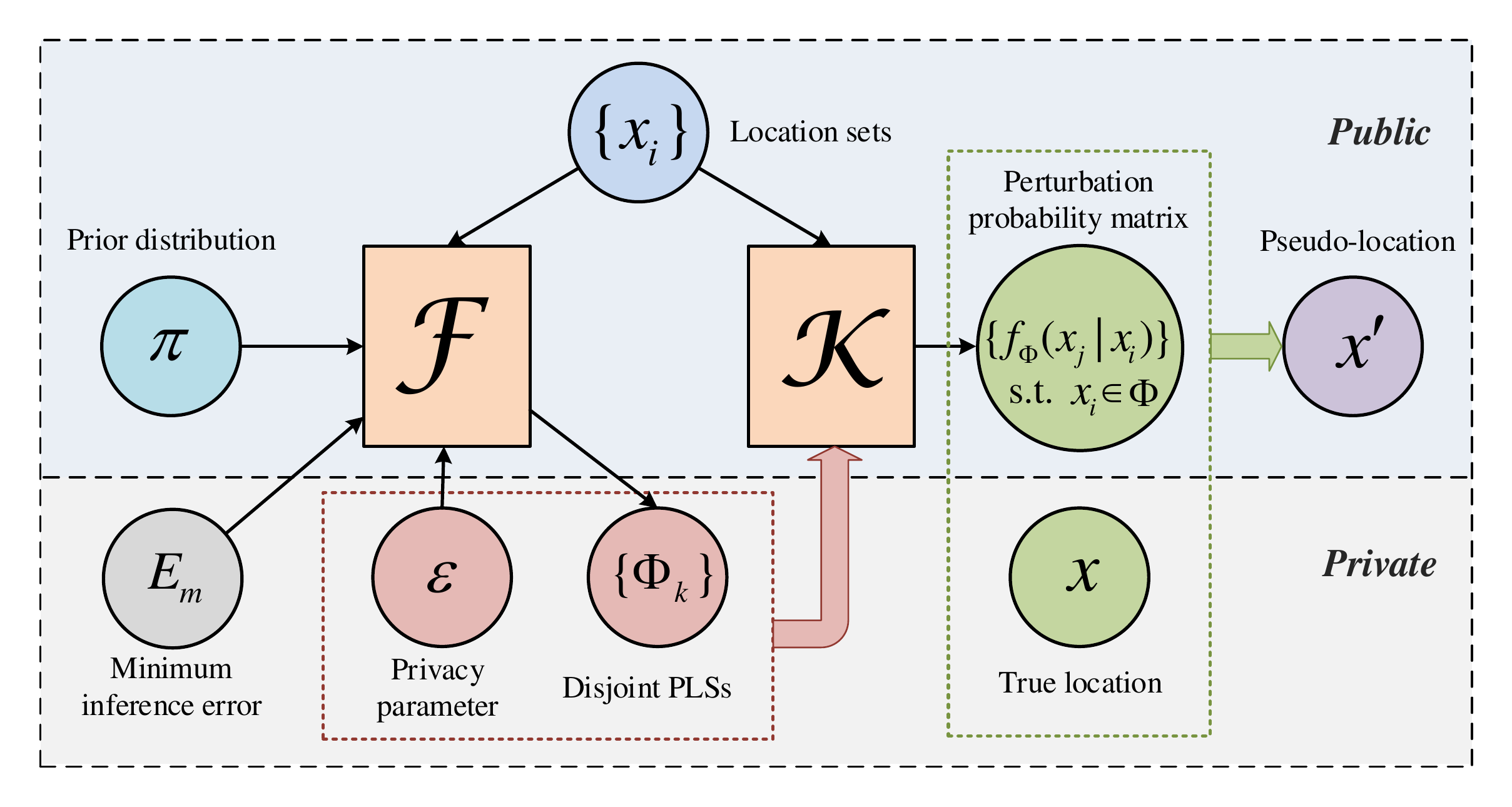}
\caption{Framework of personalized sensitivity approach.}
\label{fig:DPIVE_model}
\end{figure}

Fig.~\ref{fig:DPIVE_model} shows the procedure of the personalized sensitivity approach. In contrast to Fig. \ref{fig:Dmax_PIVE}, this approach partitions the domain into disjoint PLSs $\{\Phi_k\}$, each of which satisfies \eqref{ndss31-pri} for guaranteeing the lower bound $E_m$; then, it publishes a probability distribution matrix $\{f_\Phi(x_j|x_i)\}$, where $\Phi$ is the corresponding PLS $\Phi_k$ covering the apriori location $x_i$.

Fig.~\ref{fig:DPIVE_simu} demonstrates an example of the personalized sensitivity approach for the same scenario including $8$ locations as Fig.~\ref{fig:Dmax_simu}. There are two disjoint PLSs, which have their respective mechanism sensitivities. The two dotted circles indicate that, PLS I has a small sensitivity while PLS II has a large one. The connected solid lines show the locations each PLS contains. Thus, the blue point $3$ is out of the PLS II while being included in the red dotted circle. Besides, both PLSs can have different $\epsilon$ and should share an identical $E_m$ while satisfying condition \eqref{ndss31-pri} to personalize $\epsilon$-DP in Theorem \ref{thm:DPIVE}.


From the above, we can see that
the personalized sensitivity approach achieves
$(\epsilon_g,D(\Phi))$-geo-indistinguishability within each PLS $\Phi$, that is, for $x'\in\mathcal{X}$,
\begin{equation}
\begin{split}\label{f-over-f-geo2}
&\frac{f_{\Phi}(x^\prime|x)}{f_{\Phi}(x^\prime|y)}
=\frac{\exp \left(\frac{-\epsilon d(x,x^\prime)}{2D(\Phi)}\right)}{\exp \left(\frac{-\epsilon d(y,x^\prime)}{2D(\Phi)}\right)}
 \frac{\sum_{x^\prime \in \mathcal{X}}\exp \left(\frac{-\epsilon d(y,x^\prime)}{2D(\Phi)}\right)}{\sum_{x^\prime \in \mathcal{X}}\exp \left(\frac{-\epsilon d(x,x^\prime)}{2D(\Phi)}\right)}
\\
\leq& e^{\frac{\epsilon |d(x,x^\prime)-d(y,x^\prime)|}{2D(\Phi)}} \frac{\sum_{x^\prime \in \mathcal{X}}\exp \left(\frac{-\epsilon (d(x,x^\prime)-D(\Phi))}{2D(\Phi)}\right)}{\sum_{x^\prime \in \mathcal{X}}\exp \left(\frac{-\epsilon d(x,x^\prime)}{2D(\Phi)}\right)}\\
\leq& e^{\epsilon_g d(x,y)} e^{{\epsilon}/{2}}=e^{\epsilon_g (d(x,y)+D(\Phi))}
\le e^{2 \epsilon_g D(\Phi)}=e^\epsilon,
\end{split}
\end{equation}
where $\epsilon_g=\epsilon/(2D(\Phi))$.
It also preserves $\epsilon$-DP on each PLS while preserving $\frac{\epsilon}{2}(D(\mathcal{X})/D(\Phi_i) + D(\mathcal{X})/D(\Phi_j))$-DP on each form $\Phi_i\cup\Phi_j$,
 shortly $(\epsilon D(\mathcal{X})/D_{\min})$-DP on the whole domain $\mathcal{X}$, where $D_{\min}=\min_k D(\Phi_k)$. Indeed, for $x'\in\mathcal{X}$, and $x\in\Phi_i$, $y \in \Phi_j$, with respective sensitivities,
\begin{equation}
\begin{split}\label{f-over-f-geo-diffPLS}
&\frac{f(x'|x)}{f(x'|y)}
=\frac{\exp\left(\frac{-\epsilon d(x,x')}{2D(\Phi_i)}\right)}{\sum\limits_{s\in\mathcal{X}}\exp\left(\frac{-\epsilon d(x,s)}{2D(\Phi_i)}\right)}\cdot
\frac{\sum\limits_{t\in\mathcal{X}}\exp\left(\frac{-\epsilon d(y,t)}{2D(\Phi_j)}\right)}{\exp\left(\frac{-\epsilon d(y,x')}{2D(\Phi_j)}\right)}\\
=&\frac{\sum\limits_{t\in\mathcal{X}}\exp\left(\frac{\epsilon}{2D(\Phi_j)}\left(d(y,x')- d(y,t)\right)\right)}
{\sum\limits_{s\in\mathcal{X}}\exp\left(\frac{\epsilon}{2D(\Phi_i)}\left(d(x,x')- d(x,s)\right)\right)}\\
\leq&\exp\left(\frac{\epsilon}{2}\left(\frac{D(\mathcal{X})}{D(\Phi_j)}+\frac{D(\mathcal{X})}{D(\Phi_i)}\right)\right)
\leq \exp \left(\frac{D(\mathcal{X})}{D_{\min}}\epsilon\right).
\end{split}
\end{equation}
Similarly, this implies $(\epsilon D(\mathcal{X})/D_{\max})$-DP on $\mathcal{X}$ for the uniform sensitivity appoach.

\begin{figure}[tb]
\centering
	\subfigure[Privacy V.S. $\epsilon$]{
		\includegraphics[height=1.25 in]{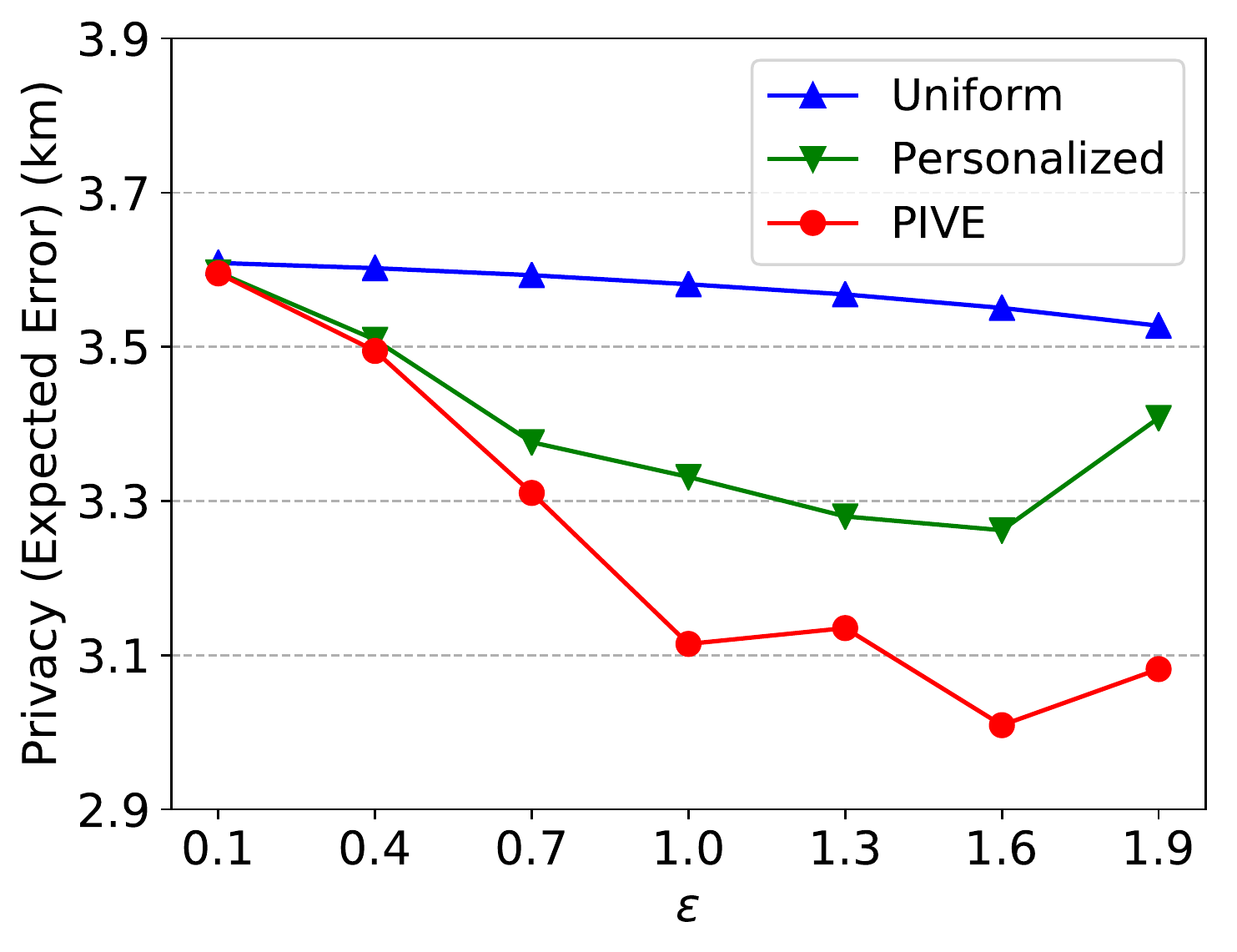}
	}
	\subfigure[Quality Loss V.S. $\epsilon$]{
		\includegraphics[height=1.25 in]{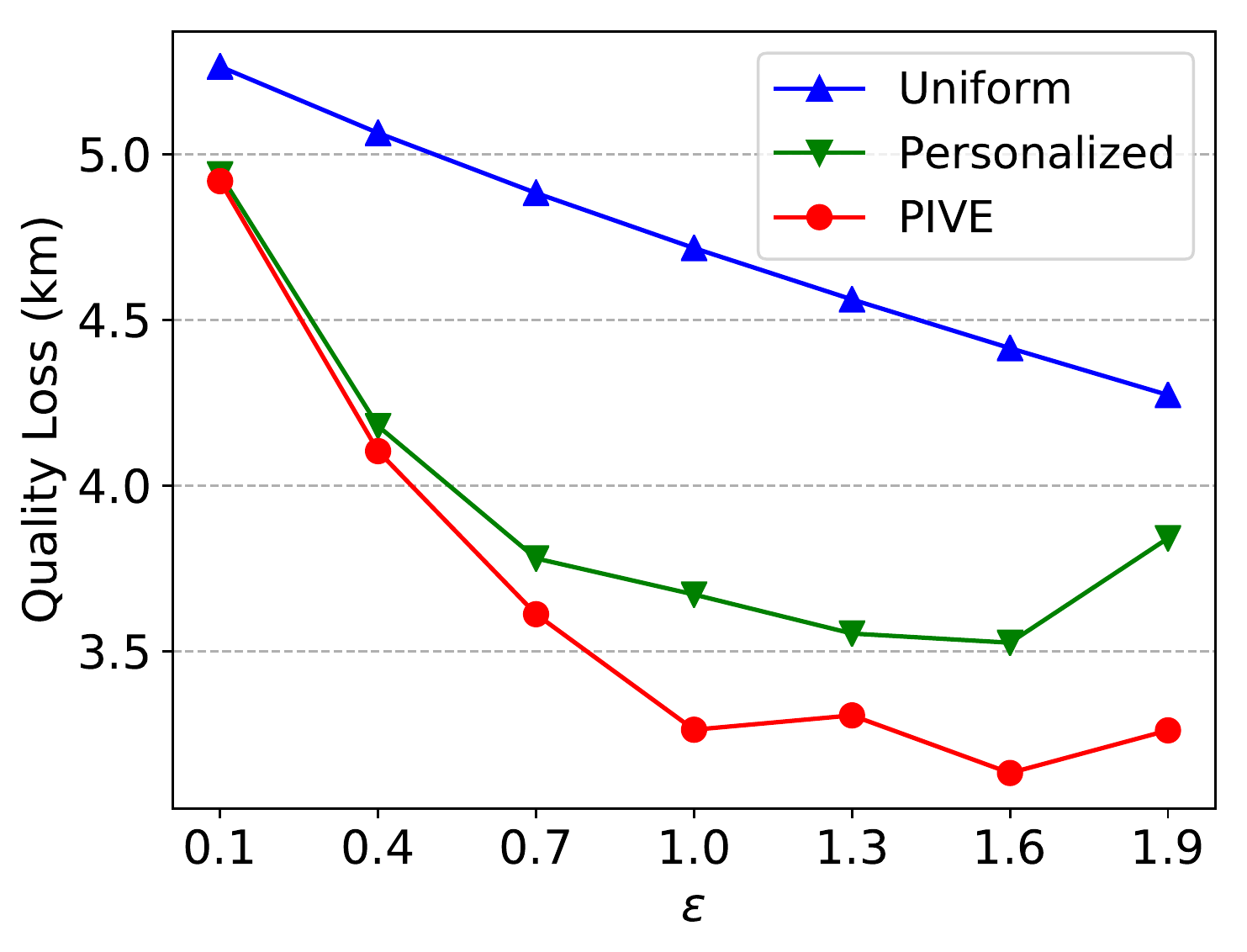}
	}
	\caption{Impact of $\epsilon$ under three strategies.}
	\label{fig:ALL}
\end{figure}

While Theorem \ref{thm:DPIVE} can be extended to allow customizable privacy demand $\epsilon_k$ over each PLS $\Phi_k$,
the current approach achieves personalized local DP at the PLS level. It provides a good tradeoff in term of data utility between the uniform sensitivity approach, which does not achieve personalization of local DP, and the PIVE, which aims (but fails) to achieve personalization by adaptively determining PLSs according to this work.
Following the setting before, Fig. \ref{fig:ALL} shows that the personalized approach with splitting for PLSs along Hilbert curve brings only $7.4\%$ growth on average  quality loss compared with PIVE and reduces $17.1\%$ quality loss on the uniform approach
while all schemes
are adjusted to satisfy \eqref{ndss31-pri}.

We conclude also the features of the personalized sensitivity approach as follows: 1) satisfying the personalization of sensitivity and privacy control knob $\epsilon$ with respect to different PLSs, and improving the data utility, 2) supporting user-defined threshold $E_m$ uniformly on PLSs, 3) achieving the $\epsilon$-DP within each PLS and weak DP on the whole domain,  4) allowing for the scenes of nonuniform location distribution or with isolated locations, and 5) not assuming the adversary's narrowing  guesses to PLSs.

\section{Related Work}\label{sec:related-work}

The growing popularity of location-based services, bringing great convenience for users,
has started raising critical privacy concerns. Many location privacy-preserving mechanisms (LPPMs) have been designed to protect users' privacy \cite{CEPP17,TS18}. 
In particular, due to an extension of Differential Privacy (DP) to arbitrary domains \cite{CAB13}, the location privacy notion, geo-indistinguishability, is introduced in \cite{ABC13} based on DP \cite{DMN06}. It ensures that two neighboring locations have similar probabilities to produce a
certain reported location. Related studies have been extended in many ways.

Usually location privacy is measured by
a function of the distance between real and inferred locations. The function can quantify
 the correctness of the adversary's inference using Hamming or Euclidean distance, as well as the
uncertainty of the adversary regarding user's location using entropy \cite{OTP17}.
While Bayesian and geo-indistinguishability approaches are combined in \cite{Sho15} with involving prior knowledge about user's locations, Oya et al. \cite{OTP17} demonstrate
that, the remapping method \cite{CEP17}, as an enhancement to
geo-indistinguishability, is helpful to improve the utility and can be regarded as a generic method to build optimal
obfuscation mechanisms in terms of average adversarial
error.
Later, Romanelli et al. \cite{RCP20} propose a machine-learning approach based on adversarial nets to
generate location obfuscation mechanisms with a good privacy utility
tradeoff.

On the applications (e.g., in crowdsourcing data publishing), Boukoros et al. \cite{BHK19} propose novel privacy and utility metrics for
evaluating the performance of LPPMs including the type of spatial obfuscation with geo-indistinguishability.
Wang et al. \cite{WYH19} first introduce differential and distortion geo-obfuscation jointly
to  task allocation for
crowdsourcing.

Yu et al. \cite{YLP17} investigate privacy leakage under Bayesian inference attacks and devise a dynamic differential location privacy
mechanism with personalized error bounds named PIVE. Their experimental evaluations show that PIVE guarantees the two privacy notions and can resist various inference attacks, particularly in the presence of skewed locations. As claimed in this paper, PIVE is representative
in geo-indistinguishability-based mechanisms though problematic in privacy preservation. This motivates the correction of PIVE in feasible directions for achieving both notions on location privacy with inference error bounds.


\section{Conclusion}\label{sec:conclusion}
In this paper, we analyze the privacy of PIVE two-phase dynamic differential location privacy
framework \cite{YLP17} and show that PIVE can not provide either provable preservation of geo-indistinguishability and distortion privacy on dynamic Protection Location Set (PLS). We demonstrate that different locations have different diameters of PLSs with intersections due to the adaptive search algorithm while the sufficient condition introduced can not control expected inference errors. Moreover, we make detailed discussions in two ways for effective mechanisms satisfying the desired privacy preferences. Both approaches proposed achieve $(\epsilon_g,\theta)$-geo-indistinguishability and local Differential Privacy (DP) on each PLS, with extensions to more general cases. The personalized sensitivity approach achieves inference error bounds without adversary's narrowing guesses to PLSs, for which we can further consider the personalization of the local DP parameter $\epsilon$ over disjoint PLSs via optimized partitions in the $2$-D space, see another work \cite{ZDC21} for details.


\vspace{-3mm}
\bibliographystyle{IEEEtran}
\bibliography{CommentPIVE}

\vskip -2\baselineskip plus -1fil
\begin{IEEEbiography}[{\includegraphics[width=1in,height=1.25in,clip,keepaspectratio]{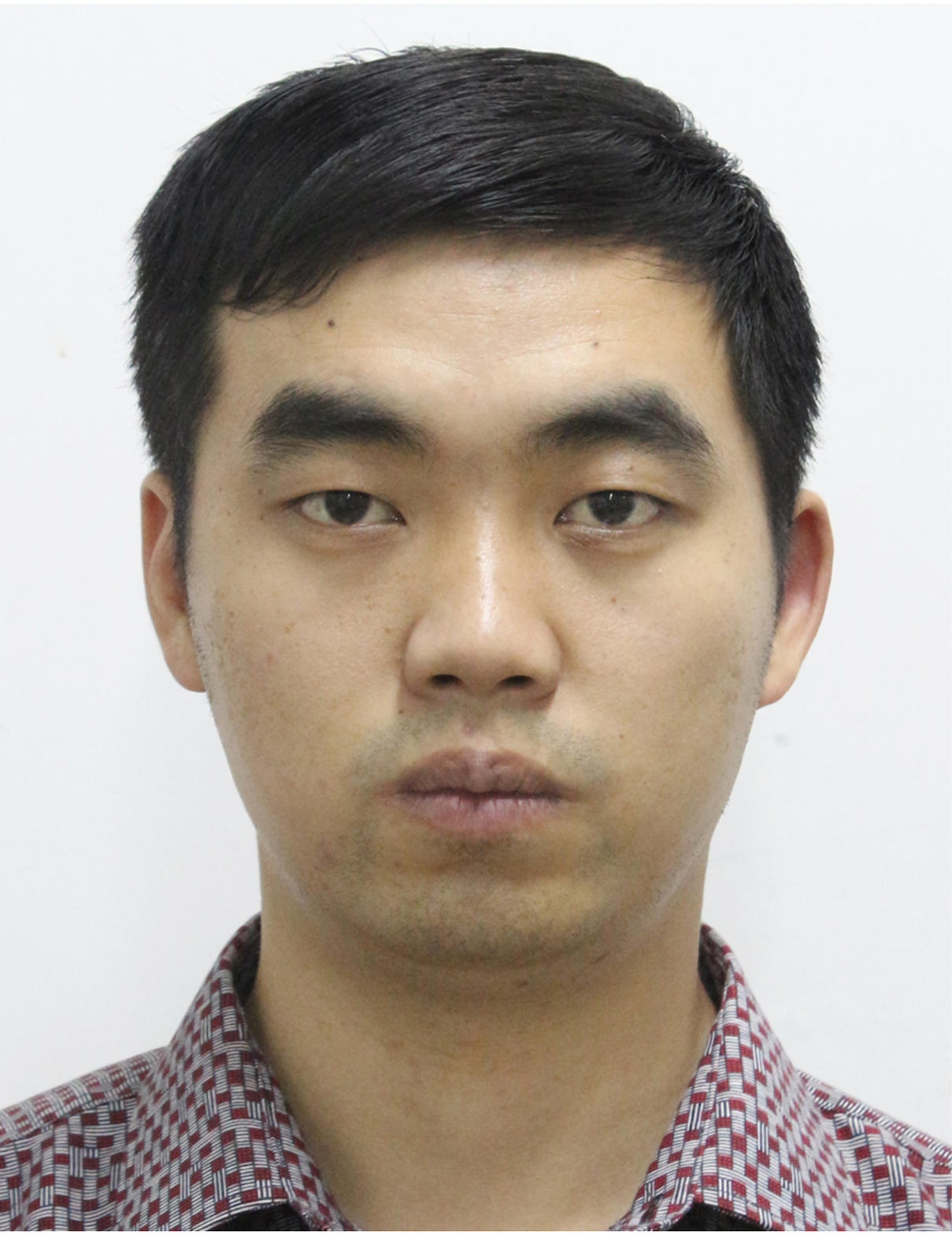}}]{Shun Zhang}
was born in Anhui Province, China, in  1982. He received his PhD degree
in applied mathematics from Beijing Normal University in 2012. He was a visiting scholar at Friedrich-Schiller-Universitat Jena, Germany, from 2014 to 2015. He is currently an associate professor at Anhui University. He has published more than 30 papers. His research interests include privacy preservation and computational complexity.
\end{IEEEbiography}

\vskip -2\baselineskip plus -1fil
\begin{IEEEbiography}[{\includegraphics[width=1in,height=1.25in,clip,keepaspectratio]{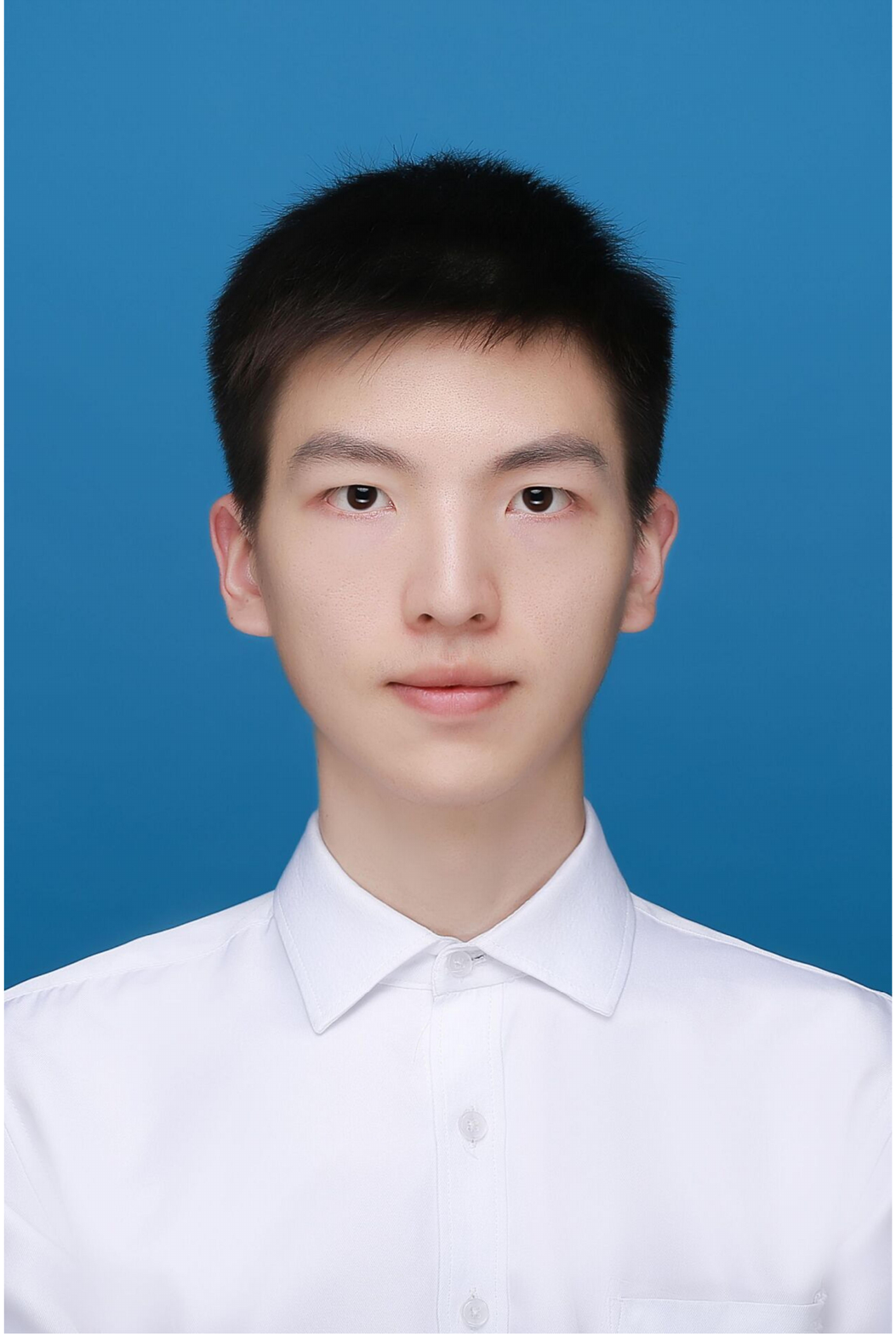}}]{Benfei Duan}
was born in Anhui Province, China, in 1997. He is currently a master student in Anhui University. His main research interests include differential privacy and location privacy.
\end{IEEEbiography}

\vskip -2\baselineskip plus -1fil
\begin{IEEEbiography}[{\includegraphics[width=1in,height=1.25in,clip,keepaspectratio]{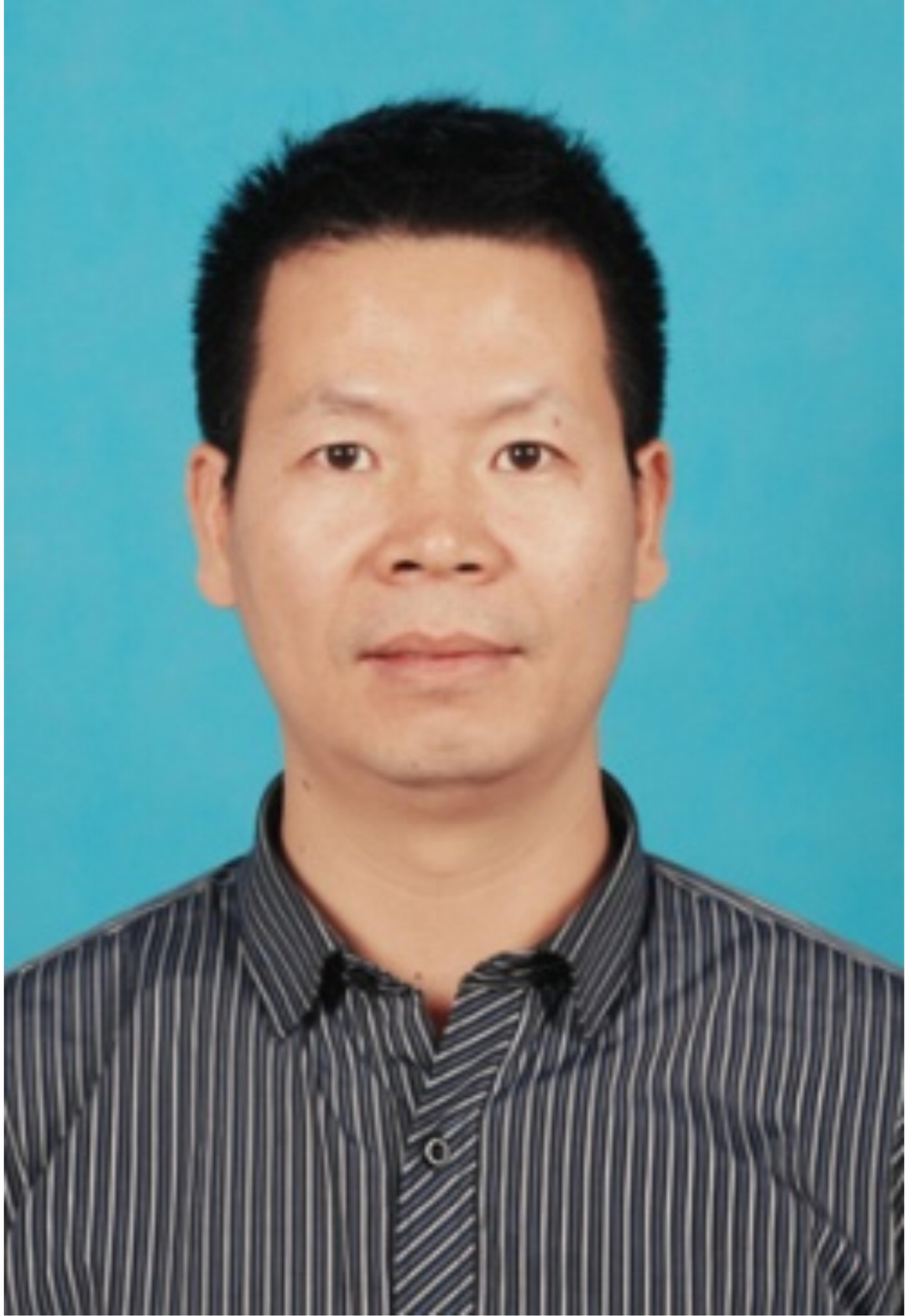}}]{Zhili Chen}
was born in Fujian Province, China, in 1980. He received his PhD degree in computer science from University of Science and Technology of China in 2009. He is currently a professor and Ph.D. supervisor at East China Normal University. He has published more than 40 papers. His main research interests include privacy preservation, secure multiparty computation, information hiding and spectrum auction. 
\end{IEEEbiography}

\vskip -2\baselineskip plus -1fil
\begin{IEEEbiography}[{\includegraphics[width=1in,height=1.25in,clip,keepaspectratio]{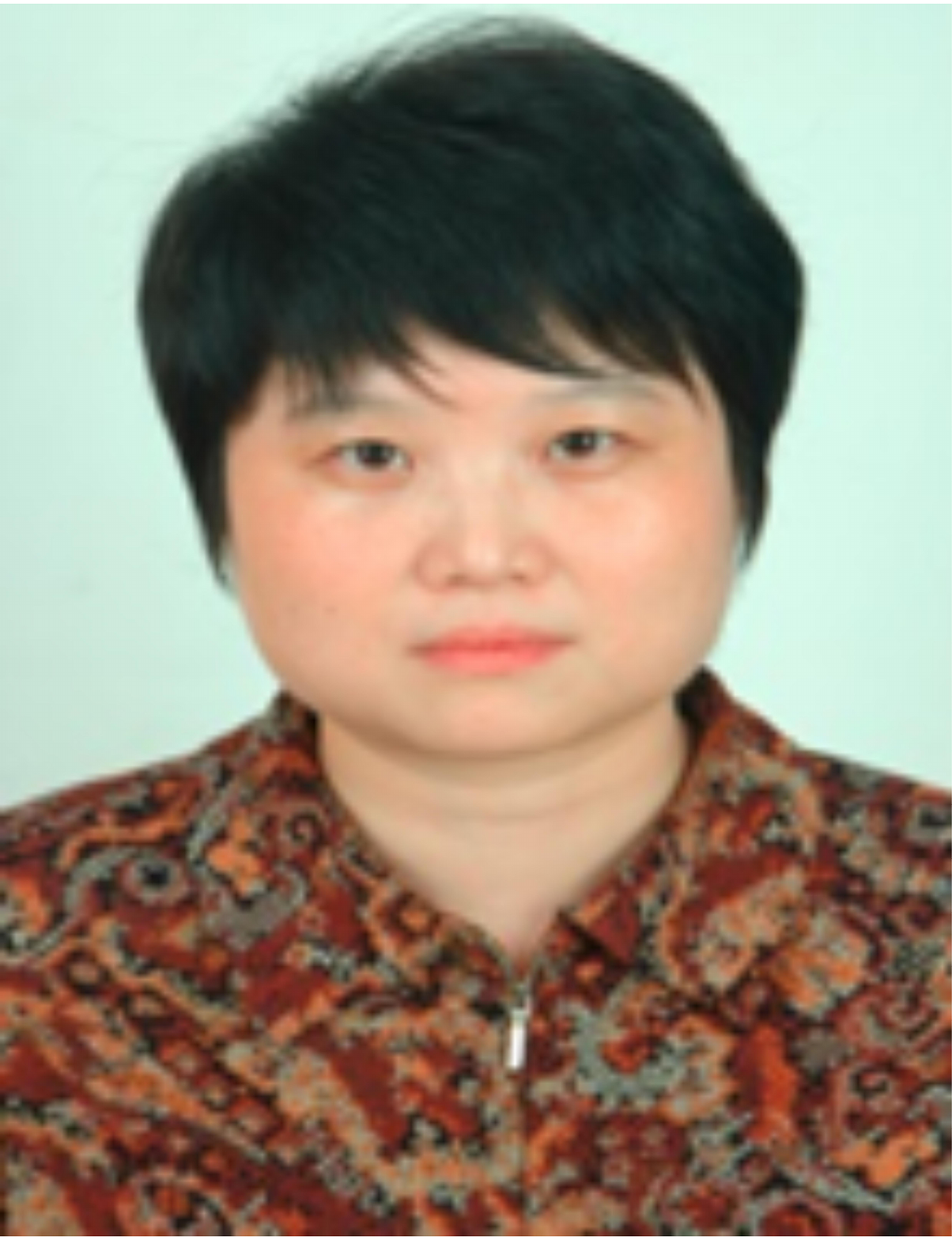}}]{Hong Zhong}
was born in Anhui Province, China, in 1965. She received her PhD degree in computer science from University of Science and Technology of China in 2005. She is currently a professor
at Anhui University. She has published more than 100 papers. Her research interests include applied cryptography, IoT security, vehicular ad hoc network, and software-defined networking (SDN).
\end{IEEEbiography}

\end{document}